 \documentstyle[aps,pre,eqsecnum,epsf]{revtex}
 
\begin{document}
 \renewcommand{\thesection}{\arabic{section}}
\twocolumn[{
 \draft
 \title{ Exact  Resummations in the Theory of Hydrodynamic Turbulence:\\
        I. The Ball of Locality and Normal Scaling}
 \author {Victor L'vov\cite{lvov}  and Itamar  Procaccia\cite{procaccia} }
 \address{Departments of~~\cite{lvov}Physics of Complex Systems
 {\rm and}~~\cite{procaccia}Chemical Physics,\\
  The Weizmann Institute of Science,
 Rehovot 76100, Israel,\\
 \cite{lvov}Institute of Automation and Electrometry,
  Ac. Sci. of Russia, 630090, Novosibirsk, Russia
   }
 \maketitle
 \widetext
 \begin{abstract}
 \leftskip 54.8pt
 \rightskip 54.8pt
 This paper is the first in a series of three papers that aim at
 understanding the scaling behaviour of hydrodynamic turbulence.  We
 present in this paper a perturbative theory for the structure functions
 and the response functions of the hydrodynamic velocity field in real
 space and time. Starting from the Navier-Stokes equations (at high
 Reynolds number Re) we show that the standard perturbative expansions
 that suffer from infra-red divergences can be exactly resummed using the
 Belinicher-L'vov transformation. After this exact (partial) resummation
 it is proven that the resulting perturbation theory is free of
 divergences, both in large and in small spatial separations.  The
 hydrodynamic response and the correlations have contributions that arise
 from mediated interactions which take place at some space- time
 coordinates.  It is shown that the main contribution arises when these
 coordinates lie within a shell of a ``ball of locality" that is defined
 and discussed. We argue that the real space-time formalism developed
 here offers a clear and intuitive understanding of every diagram in the
 theory, and of every element in the diagrams. One major consequence of
 this theory is that none of the familiar perturbative mechanisms may ruin
 the classical Kolmogorov (K41) scaling solution for the structure
 functions. Accordingly, corrections to the K41 solutions should be sought
 in nonperturbative effects.  These effects are the subjects of papers II
 and III in this series, that will propose a mechanism for anomalous
 scaling in turbulence, which in particular allows multiscaling of the
 structure functions.
  \end{abstract}
 \leftskip 54.8pt
}]
 \narrowtext
 \section{Introduction to the series of papers}

  This paper is the first in a series of three papers that aim for a
  systematic elucidation of the analytic structure of the the statistical
 theory of hydrodynamic turbulence at high Reynolds number Re.  The
 natural quantities that appear in the statistical description of
 turbulence are averages of products of hydrodynamic fields at different
 space-time points. The aim of the analytic approach is to evaluate such
 quantities on the basis of the equations of fluid mechanics.

 The fundamental field in hydrodynamics is the velocity field of the
 fluid, denoted as ${\bf u}({\bf r},t)$ where $\bf r$ is a point in
 $d$-dimensional space (usually $d=2$ or 3) and $t$ is the time.
 Statistical quantities that attracted years of experimental and
 theoretical attention \cite{MY,91Sre,94Nel} are the structure functions
 of velocity differences, denoted as $S_n(R)$
  \begin{equation}
 S_n(R) = \left\langle\vert\hbox{\bf u}({\bf
 r}+{\bf R},t) -{\bf u}({\bf r},t)\vert^n\right\rangle \ .
 \label{a5}
 \end{equation}
 where $\left<\dots\right>$ stands for a suitably defined ensemble
 average.  It has been stipulated for a long time that the structure
 functions scale as a function of $R$ when $R$ is in the so called
 ``inertial range", i.e.  $\eta \ll R \ll L$ with $\eta$  being the inner
 viscous scale and $L$ being the outer integral scale of turbulence:
 \begin{equation}
 S_n(R) \sim R^{\zeta_n} \,, \quad
 \eta \ll R \ll L 		\label{scalelaw}
 \end{equation}
 with $\zeta_n$ being scaling exponents. One of the major questions in
 fundamental turbulence research is whether these scaling exponents are
 correctly predicted by the classical Kolmogorov 41 theory in which
 $\zeta_n = n/3$, or whether these exponents manifest the phenomenon of
 ``multiscaling" with $\zeta_n$ being a nonlinear function of $n$, as has
 been indicated by experiments.

 Experimental research did not confine itself to the measurement of the
 structure functions of velocity differences. Gradient fields featured  as
 well.  For example the correlation function of the energy dissipation
 field has been studied extensively. The dissipation field $\epsilon({\bf
 r},t)$ is defined as
  \begin{equation}
 \epsilon({\bf r},t) \equiv {\nu\over 2}[\nabla_\alpha u_\beta({\bf r},t)+
 \nabla_\beta u_\alpha({\bf r},t)]^2    \label{defeps}
 \end{equation}
 with $\nu$  being the kinematic viscosity. The correlation function of the
 dissipation field $K_{\epsilon\epsilon}(R)$ is
 \begin{equation}
 K_{\epsilon\epsilon}(R) = \left< \hat\epsilon ({\bf r}+{\bf R},t)
 \hat\epsilon ({\bf r},t) \right> \ , \label{Kee}
 \end{equation}
 where  $\hat\epsilon({\bf r},t) = \epsilon ({\bf r},t) - \bar\epsilon$.
 Here and below $\bar\epsilon\equiv \left<\epsilon\right>$.  It was
 claimed experimentally \cite{MY,91Sre,94Nel} that
 $K_{\epsilon\epsilon}(R)$ decays as a power law,
  \begin{equation}
 K_{\epsilon\epsilon}(R) \sim  R^{-\mu} \,,\quad 	\eta \ll  R \ll L
\label{Rmiu}
 \end{equation}
 with $\mu$ having a numerical value of $0.2-0.3$. The analytic derivation
 of this law from the equations of fluid mechanics, and the calculation of
 the numerical value of the scaling exponent $\mu$  have also been elusive
 goals of the theoretical research.

 In the present series of three papers we will present a systematic theory
 that culminates in the analytic estimate of the scaling exponents
 $\zeta_n$ and $\mu$, in addition to the calculation of other quantities
 of experimental and theoretical interest.  Since the road to these
 results is not straightforward, we present here a summary of the three
 papers in the series, hoping to motivate the reader to examine the
 necessary steps of analysis even though they are somewhat tedious. We
 believe that the proposed structure of the resulting scaling theory is
 rather beautiful and simple, justifying in part the lengthy path taken to
 reach it.

 The present paper which is numbered I deals with the perturbative theory
 of the correlation, response and structure functions of hydrodynamic
 turbulence.  The main result of this paper (and see also \cite{87BL}) is
 that after appropriate resummations and renormalizations the perturbative
 theory for these quantities is finite order by order in the sense that
 all the integrals appearing in the theory are convergent in the
 ultraviolet and the infrared limits . The meaning of this result is that
 there is no perturbative mechanism to introduce a length scale into the
 theory of the structure functions. When the perturbative series diverges
 either in the infrared or in the ultraviolet the cutoff lengths that are
 used to tame the divergence serve as  renormalization scales which can be
 used to change the scaling exponents from the prediction of dimensional
 analysis. This result indicates that as far as the perturbative theory is
 concerned there is no mechanism to correct the exponents $\zeta_n$ from
 their K41 values. Of course, nonperturbative effects can very well
 furnish such a mechanism and therefore in paper II \cite{95LP-b}
 (subtitled "A Ladder to Anomalous Scaling) we turn to the analysis of
 nonperturbative effects. There is more than one source of nonperturbative
 effects that need to be studied. One is obtained from summing up infinite
 perturbative series into analytic equations that display solutions that
 are not obtainable from the initial perturbation series. Another one
 stems from sum rules or constraints on the perturbation theory as a
 whole. In paper II  we study infinite resummations of ladder type
 diagrams. It is shown there that the renormalized perturbation theory for
 correlation functions that include velocity  derivatives (to second or
 higher power) exhibit in their perturbation expansion a logarithmic
 dependence on the  viscous scale $\eta$ \cite{95LL,94LL}. In this way the
 inner scale of turbulence appears explicitly in the analytic theory. The
 perturbative series can be resummed to obtain integrodifferential
 equations for some many-point objects of the theory. These equations have
 also non-perturbative scale-invariant solutions that can be represented
 as power laws of $\eta$ to some exponents $\Delta$. For example it will
 be shown in paper III \cite{95LP-c} subtitled (Scenarios for Anomalous
 Scaling and Intermittency) that the correlation function of the energy
 dissipation field has such dependence:
  \begin{equation}
 K_{\epsilon\epsilon}(R) \sim \bar
 \epsilon^2 (L/R)^{2\zeta_2-\zeta_4} (R/\eta)^
 {2(\Delta-\Delta_c)}
 \label{Kee1}
 \end{equation}
 where $\Delta_c=2 - \zeta_2$. It has been argued \cite{95LP} that if
 $\Delta<\Delta_c$ (a situation referred to as the ``subcritical
 scenario"), then K41 is asymptotically exact for infinite Re. Then
 $2\zeta_2=\zeta_4 = 4/3$ and the outer scale $L$ disappears from
 (\ref{Kee1}). In that case the exponent $\mu$ is identified with
 $2(4/3-\Delta)$, and the renormalization length is the inner length
 $\eta$. In fact, in paper II it will be shown that the exponent $\Delta$
 can be computed exactly, and that it takes on exactly the critical value
 $\Delta = \Delta_c$. As a result of this the correlation
 $K_{\epsilon\epsilon}(R)$ can be shown to depend on $R$ like
 \begin{equation}
 K_{\epsilon\epsilon}(R) \simeq
  \bar\epsilon^2 (R/L )^{\zeta_4-2\zeta_2} \ .\label{Kee2}
 \end{equation}
 In other words, the critical situation $\Delta = \Delta_c$ results in the
 disappearance of the inner renormalization scale and the appearance of
 the outer renormalization scale in (\ref{Kee2}). In addition one notes
 that $K_{\epsilon\epsilon}(R)$ decays as a function of $R$ (i.e. the
 correlation is mixing) only if $\zeta_4 < 2\zeta_2$  which implies
 deviations from K41. Thus we will argue in paper II that the critical
 scenario $\Delta = \Delta_c$ goes hand in hand with multiscaling if
 $K_{\epsilon\epsilon}(R)$ is mixing and then $\mu$ is identified with
 $\zeta_4 - 2\zeta_2$.  Paper III which is the last paper in this series
 focuses on the analysis of the nonperturbative constraint furnished by
the ballance equation which is obtained from the equations of motion of
the structure functions. This analysis exposes a mechanism for
multiscaling which leads to the evaluation of the scaling exponents
 $\zeta_n$ from first principles.

 \section{On the background of the analytic description of turbulence}

 Of foremost importance in the analytical approach have been those methods
 that rely on some perturbative treatment of the Navier-Stokes
 equation\cite{61Wyl,73MSR,77Kra,76Dom,76Jan,87BL}. Such treatments are
 fraught with difficulties, which stem basically from two sources. The
 main source is the fundamental fact that turbulence has no small
 parameter. Thus, any naive perturbation scheme is bound to fail, being
 effectively a non-convergent series in powers of Re. Any truncation of
 such a series leads to disastrous nonsense. Evidently one must cleverly
 resum such series to obtain a formal perturbation expansion in terms of
 an effective coupling constant which differs from Re. The second source
 of difficulty is that the natural statistical objects which appear in the
 perturbation expansions\cite{61Wyl,73MSR,76Dom,76Jan} of the
 Navier-Stokes equation are correlation functions ${\cal
 F}_{\alpha\beta}({\bf r},{\bf r}',t,t')$ of the velocity field ${\bf
 u}({\bf r},t)$ itself:
 \begin{equation}
 {\cal F}_{\alpha\beta}({\bf r},{\bf r}',t,t')
 = \left\langle u_\alpha({\bf r},t)
 u_\beta({\bf r}',t')\right\rangle \ .
 \label{a1}
 \end{equation}
 The problem is that the correlator
 $F_{\alpha\beta}({\bf r},{\bf r}',t,t')$ is not universal, since it is
 dominated by contributions to $u_\alpha({\bf r},t)$ which come from the
 largest scales in the fluid flow which are determined by the features of
 the energy injection mechanisms. This physical fact is reflected in the
 theory as infra-red divergences that have plagued the discussion of  the
 analytic approach for decades. Indeed, all the early attempts to  develop
 a consistent analytic approach to turbulence, notably the well  known
 theories of Wyld\cite{61Wyl} and of Martin, Siggia and Rose\cite{73MSR},
 share this problem.

 It is ironic that the appropriate universal objects were known all along
 due to the insight of Richardson\cite{22Ric,26Ric} and later of
 Kolmogorov\cite{41Kol}, who realized that velocity {\it differences}
 across a lengthscale $R$ do possess good properties, since they are
 dominated by motions of scale $R$. In Kolmogorov's language such a
 velocity difference is expected to be universal when $R$ is in ``inertial
 range".  It seems advisable to develop the analytic techniques also in
 terms of similar universal objects.  To this aim one may transform the
 equations of motion to new variables.  A potentially promising approach
 could be to recast the Navier-Stokes equation in terms of Lagrangian
 coordinates as has been attempted by Kraichnan\cite{77Kra}. The equation
 for the Lagrangian velocity involves however the Eulerian velocity which
 relates to the Lagrangian velocity via a complicated time evolution
 operator. A lot has been accomplished in this approach; firstly,
Kraichnan acheived a qualitative undertsanding of the different type of
power law spectra that appear in different contexts like scalar pasive
diffusion, two and three dimensional turbulence, and Burgers dynamics.
Secnodly, a numerical prediction for the Kolmogorov constant was obtained.
However, the resulting perturbation theory does not have a diagrammatic
 representation in the sense that one does not know how to achieve
 infinite resummations.  The progress in understanding non-perturbative
aspects remained unsatisfactory.

 Another approach that claimed success was due to Belinicher and
 L'vov\cite{87BL} who introduced an exact transformation to variables
 which they called ``quasi-Lagrangian". This term was not particularly
 successful; on the one hand it could give the impression of an approximate
 scheme, and on the other hand it implied proximity to the Lagrangian
 transformation. Both impressions are incorrect. The Belinicher-L'vov
 transformation is exact, and it has very little to do with the cumbersome
 features of the Lagrangian
 transformation. In terms of the Eulerian velocity ${\bf u}({\bf r},t)$
 they defined the field ${\bf v}({\bf r}_0,t_0\vert {\bf r},t)$ as
 \begin{equation}
 {\bf v}({\bf r}_0,t_0\vert {\bf r},t)\equiv {\bf u}\lbrack{\bf r}
 +\mbox{\boldmath$\rho$}
 ({\bf r}_0,t),t\rbrack
 \label{a2}
 \end{equation}
 where
 \begin{equation}
 \mbox{\boldmath$\rho$}  ({\bf r}_0,t)
  =\int_{t_0}^{t}{\bf u }[{\bf r}_0 +\mbox{\boldmath$\rho$}({\bf
  r}_0,\tau) ,\tau] \ .
  \label{a3}
 \end{equation}
 Note that $\mbox{\boldmath$\rho$}  ({\bf r}_0,t)$ is precisely the
 Lagrangian trajectory of a fluid particle that is positioned at ${\bf
 r}_0$ at time $t=t_0$. Nevertheless the field  ${\bf v}({\bf r}_0,t_0\vert
 {\bf r},t)$ becomes in time very different from the Lagrangian field in
 which the Lagrangian trajectories of all space points are involved.  The
 observation of Belinicher and L'vov was that the variables ${\bf
 w}(\hbox{\bf r}_0,t_0\vert \hbox{\bf r},t)$ which are defined as
 \begin{equation}
 \hbox{\bf w}(\hbox{\bf r}_0,t_0\vert \hbox{\bf r},\hbox{\bf t})
 \equiv{\bf v}(\hbox{\bf r}_0,t_0\vert \hbox{\bf r},t) -{\bf v}
 (\hbox{\bf r}_0,t_0\vert \hbox{\bf r}_0,t)
 \label{a4}
 \end{equation}
 exactly satisfy the Navier-Stokes equation in the limit of incompressible
 fluid.  Consequently they could develop a diagrammatic perturbation theory
 in terms of these variables. The resulting theory was free of the two
 related problems that we discussed above: the $\left\langle{\bf w}{\bf
 w}\right\rangle$ correlators were universal for $\vert{\bf r}-{\bf
 r}_0\vert$ in the inertial range, and the theory was free of infra-red
 (and ultra-violet) divergences. They also showed that the Kolmogorov 1941
 scaling is an order by order solution of the resulting theory. Two
 important  properties of this formulation are: (i) The simultaneous
 correlators of ${\bf v}\lbrack\hbox{\bf r}_0\vert\hbox{\bf
 r},t\rbrack$ are identical to the simultaneous correlators of $\hbox{\bf
 u}(\hbox{\bf r},t)$. The  reason is that for stationary statistics the
 simultaneous correlators of  an arbitrary number of factors of ${\bf
 v}\lbrack\hbox{\bf r}_0\vert\hbox{\bf r},t\rbrack$ does not depend on $t$,
 and in particular one can take $t=t_0$. The property then follows directly
 from Eq.(\ref{a3}). (ii) The correlators of ${\bf w}$ are closely
 related to the structure functions of {\bf u} (\ref{a5}).
 Clearly
 \begin{equation} S_n(\vert\hbox{\bf
 r}-\hbox{\bf r}_0\vert) = \left\langle\vert{\bf w}(\hbox{\bf
 r}_0,t_0\vert\hbox{\bf r},t)\vert^n\right\rangle \ .
 \label{a6}
 \end{equation}
 As the structure functions of the Eulerian velocity differences were for
 years at the focus of experimental research, the formulation in terms  of
 the variables ${\bf w}$ gives a direct link between theory and
 experiments.  We will refer to the variable ${\bf w}(\hbox{\bf
 r}_0,t_0\vert\hbox{\bf r},t)$ as the BL velocity differences.

 For some reason these results have not been widely accepted.  One reason
 must have been the fact that the theory involved the subtraction of a
 velocity field at {\it one special point} (denoted $\hbox{\bf r}_0$
 above),  and this, in the words of Kraichnan\cite{94Kra-a}, ``cannot cure
 the infra red  divergences that stem from far away regions". Possibly the
 main reason for disbelief is that the Kolmogorov solution was found to be
 an exact order by order solution of the theory. There are two reasons not
 to believe a theory that reproduces K41. Experimentally, it was widely
 stated that observations of the structure functions of velocity
 differences are in contradiction with the  K41 scaling,
 see for example \cite{93SS,93BCTBMS} and references therein.
 Theoretically, it was believed since the days of Landau that the strong
 dissipation fluctuations ``must" renormalize the exponents that are found
 within ``naive" dimensional analysis (as is commonly found in the context
 of critical phenomena etc.). Lastly, the issue of the vertex
 renormalization was left out in \cite{87BL} for future work. This also
 gave reasons to doubt whether the demonstration of the K41 scaling
 solution is to be accepted. The wisdom of the community about this issue
 was summarized recently in one sentence\cite{94MW}: ``The quasi-
 Lagrangian formulation serves only to provide a sophisticated diagrammatic
 reformulation of the Kolmogorov argument".

 The aim of this paper is to clarify and solidify the theoretical points
 that are needed to construct a Galilean-invariant and divergence-free
 renormalized perturbation scheme for turbulence, and to answer all the
 doubts described above. Our strategy is to develop the theory from the
 start entirely in ${\bf r},t$ representation in terms of new propagators
 which are correlations and Green's functions of the BL velocity
 differences (\ref{a4}); these objects possess the good properties of being
 universal and of having transparent physical significance. We find that
 this new approach enjoys the merit that each diagram and the elements of
 the diagrams have definite physical interpretation.  The simultaneous
 correlation functions of our theory are the functions of the Eulerian
 velocity differences. We shall see that the physical reason for the lack
 of infra-red divergences is the local character of the hydrodynamic
 interactions. This character is reflected in the fast decay of the Green's
 functions when one of its spatial arguments goes away from a ``ball of
 locality" whose radius is computed in the theory. The intuitive idea of
 the locality of interaction in fluids in ${\bf r}$ space finds an
 immediate reflection in the structure of the theory. Only interactions
 inside the ball of locality turn out to be important.

 As a result of this theory we will explain why the K41 scaling is indeed
 a consistent order by order solution of the theory. Notwithstanding, we
 will argue in papers II and III of this series that nonperturbative
 effects may furnish a multiscaling solution which differs from K41. The
 essence of the theory is that such a solution cannot be seen in order by
 order expansions.

 The structure of this paper is as follows: In Section 2 we
 present the equations of motion, introduce the appropriate variables,
 and derive, via the path integral formulation, the Dyson-Wyld
 equations for the second order correlators and Green's functions (i.e
 the propagators). In Section 3 we discuss the physical meaning of the
 propagators and of the mass operators in our formulation, study the
 properties of the propagators, and show how they depend on time and
 on the coordinates. The crucial property of the Green's function in BL
 variables is that its bare value is given as the difference between two
 projection operators. This serves as an initial value for the equation of
 motion of our Green's function, leading to solutions that differ
 fundamentally from the Green's function in the Eulerian
 representation. The properties discovered in Section 3 allow us to
 present in Section 4 the proof of locality. This is the central section of
 this paper;  it is shown that all the diagrams that appear in the infinite
 series for the mass operators in the Dyson-Wyld equations are local.
 This means that they have no divergences either in the large space-
 time separation limit or in the small separation limit, and they all
 contribute mostly in the ``ball of locality" that is defined in section 3.
 As a result we will be able to show that the only consistent scaling
 solution (order by order) has the K41 scaling exponents. Of course, this
 approach suffers from the usual problem that although the series
 converges order by order, it may diverge as a whole. This is the possible
 source of nonperturbative effects that are being discussed in papers II and
 III of this series. Section
 5 is reserved for a summary and a discussion.

 \section{Equations of Motion, Variables and  Renormalizations}
 \subsection{Equations of motion}
 The analytic theory of turbulence is based on the Navier-Stokes equation
 for the Eulerian velocity field ${\bf u}({\bf r},t)$. In the case of an
 incompressible fluid they read
 \begin{equation}
  \partial   {\bf u}/ \partial t + (   {\bf u}\cdot
  {\bbox{\nabla}})   {\bf u} -\nu\nabla^2   {\bf u}
 +{\bbox{\nabla }} p = {\bf f}, \quad {\bbox{\nabla}}\cdot   {\bf u}=0\,,
 \label{b1}
 \end{equation}
 where $\nu$ is the kinematic viscosity, $p$ is the pressure, and ${\bf f}$
 is some  forcing which maintains the flow. Since we are interested in
 incompressible flows, we project the longitudinal components out of the
 equations of motion. This is done with the help of the projection operator
 $\raisebox{.2ex}{$\stackrel{\leftrightarrow}{\bf P}$}$ which is formally
 written as
 $\raisebox{.2ex}{$\stackrel{\leftrightarrow}{\bf P}$}$
 $\equiv -\nabla^{-2}{\bbox{\nabla  }}\times\nabla\times$. The application of
 ${\raisebox{.2ex}{$\stackrel{\leftrightarrow}{\bf P}$} }$
 to any given vector field $   {\bf a}( {\bf r})$ is non local, and it has
 the form:
 \begin{equation}
 \lbrack\raisebox{.2ex}{$\stackrel{\leftrightarrow}{\bf P}$}
  {\bf a}(
 {\bf r})\rbrack_\alpha =\int d   {\bf r}'   P_{\alpha\beta}(
 {\bf r}-{\bf r}')a_\beta(   {\bf r}'),
 \label{b2}
 \end{equation}
 where $P_{\alpha\beta}(   {\bf r}-   {\bf r}')$ is the inverse Fourier
 transform in {\bf k} of the tensor $P_{\alpha\beta}(   {\bf k})$:
 \begin{eqnarray}
 P_{\alpha\beta}( {\bf r}- {\bf r}')&=&\int{d^3k\over (2\pi)^3}
 \exp\lbrack -i(   {\bf r}-   {\bf r}')\cdot {\bf k}\rbrack
 P_{\alpha\beta}(   {\bf k}),
 \label{b3} \\
 P_{\alpha\beta}( {\bf k})&=&\delta_{\alpha\beta}-{1\over k^2}k_\alpha
 k_\beta\ .
 \label{b4}
 \end{eqnarray}
  The calculation of the Fourier integral (\ref{b3}) gives
 \begin{eqnarray}
 P_{\alpha\beta}({\bf r}&-&{\bf r}')
 =\delta_{\alpha\beta}\delta({\bf r}-{\bf r}' )
 \label{b5} \\
 &-&{1\over 4\pi}\left[{\delta_{\alpha\beta}
 \over \vert{\bf r}-{\bf r}' \vert^3}
 -3{(r_\alpha-r'_\alpha)(r_\beta-r'_\beta)
 \over\vert{\bf r}-{\bf r}'\vert^5}\right] \ .
 \nonumber
 \end{eqnarray}
 Applying  {\raisebox{.2ex}{$\stackrel{\leftrightarrow}{\bf P}$} }
 to Eq.(\ref{b1}) we find
 \begin{equation}
 (\partial/ \partial t-\nu\nabla^2){\bf u}+
 {\raisebox{.2ex}{$\stackrel{\leftrightarrow}{\bf P}$} }
 ({\bf u}\cdot\bbox{\nabla}){\bf u}
 ={\raisebox{.2ex}{$\stackrel{\leftrightarrow}{\bf P}$} }\,
 {\bf f}\ .
 \label{b6}
 \end{equation}
 We now perform the Belinicher-L'vov change of variables (\ref{a2}-\ref{a3})
 together with ${\bf f}[{\bf r}
 +\mbox{\boldmath$\rho$}({\bf r}_0,t_0,t),t]
 =\tilde{\mbox{\boldmath$\phi$}}
 \lbrack{\bf r}_0,t_0\vert {\bf r},t\rbrack$.
  Using the Navier Stokes equation and the chain rule of
 differentiation we find the equation of motion for ${\bf v}\lbrack
 {\bf r}_0,t_0\vert   {\bf r},t\rbrack$:
 \begin{eqnarray}
 {\partial\over \partial t}{\bf v}
 \lbrack   {\bf r}_0,t_0\vert {\bf r},t\rbrack
 &+&\raisebox{.2ex}{$\stackrel{\leftrightarrow}{\bf P}$}
 \lbrack{\bf w}\lbrack   {\bf r}_0,t_0\vert {\bf r},t\rbrack\cdot
 \mbox{\boldmath$\nabla$}\rbrack{\bf w}\lbrack {\bf r}_0,t_0\vert
 {\bf r},t\rbrack
 \nonumber  \\
 &=&\nu[\nabla^2_0+\nabla^2]{\bf w}\lbrack   {\bf r}_0,t_0\vert
 {\bf r},t\rbrack +
 {\raisebox{.2ex}{$\stackrel{\leftrightarrow}{\bf P}$} } \,
 \mbox{\boldmath$\phi$}
 \label{b7}
 \end{eqnarray}
 where ${\bf w}( {\bf r}_0,t_0\vert   {\bf r},t ) $ is the BL velocity
 difference defined in (\ref{a4}) and $\nabla^2_0$ acts on the
 ${\bf r}_0$  variable. We note  that the equation of motion is
 independent of $t_0$, and from now on we drop the argument $t_0$ in ${\bf
 v}( {\bf r}_0\vert   {\bf r},t) $.  Our aim is to develop a theory in
 terms of the BL field ${\bf w}$. It can be seen from (\ref{b7}) that ${\bf
 w}$ satisfies the Navier-Stokes equation (with a different forcing).
 However we will use the mixed equation (\ref{b7}) since the change of
 variables from ${\bf u}$ to ${\bf v}$ is more direct.

 \subsection{Generating functional and perturbation theory}
 The development of perturbation theory can be done using a number of
 equivalent formalisms, such as the Wyld diagrammatics\cite{61Wyl} or the
 path integral formulation\cite{73MSR,76Dom,76Jan,93KL,95LP-a}. Both
 approaches have been widely reviewed, and can be shown to give identically
 the same diagrams\cite{95LP-a}. We follow here the approach of Ref.
 \cite{76Dom,76Jan,95LP-a} which is based on the generating functional
 $Z(\mbox{\boldmath$\lambda$},{\bf m})$
 \begin{eqnarray}
 Z(\mbox{\boldmath$\lambda$} ,   {\bf m})
 &=&{1\over Z}\int D\,{\bf u}(x) D\,{\bf p}(x)\, \exp\lbrace iI
 \nonumber \\
 &&+\int dx\lbrack {\bf u}(x)\cdot \mbox{\boldmath$\lambda$} (x)
 +   {\bf p}(x)\cdot   {\bf m}(x)\rbrack\rbrace\,,
 \label{b8a}
 \end{eqnarray}
 where $x\equiv({\bf r},t)$. The quantity $I$ is referred to as the
 effective action. It  is customary to divide $I$ into two parts, one
 quadratic and the other  triadic in the field {\bf p} and {\bf u}:
 \begin{eqnarray}
 I&=&I_0+I_{\rm int}\,,
 \label{b9a}  \\
 I_0&=&\int dx\lbrack p_\alpha{\partial u_\alpha\over\partial t}
 -\nu p_\alpha \nabla^2  u_\alpha\rbrack
 \label{b10a} \\
 &&+{i\over 2}\int p_\alpha (x)D_{\alpha\beta}(x-y) p_\beta (y)\,dx\,dy\,,
 \nonumber \\
 I_{\rm int}
 &=&\int dx\,p_\alpha (x) u_\beta (x) \nabla_\beta u_\alpha (x)\ .
 \label{b11a}
 \end{eqnarray}
 Here $D_{\alpha\beta}(x-y)$ is the correlation function of a Gaussian
 random force  ${\bf f}$.  We will not work in these variables, but rather
 make a change of variables in (\ref{b8a}-\ref{b11a}) according to
 (\ref{a2}-\ref{a4}) and (\ref{b7}).  We will  denote the resulting
 function as ${\cal Z}({\bf l}, {\bf m})$:  \begin{eqnarray} {\cal Z}({\bf
 l},{\bf m})& =&{1\over Z}\int D {\bf v} D   {\bf p}\,\exp\lbrace i   I
 \nonumber \\
 &&+\int dx\, {\bf v}({\bf r}_0\vert x)
 \cdot {\bf l}(x)+ {\bf p}(x)\cdot {\bf m}(x)\rbrack\rbrace\ .
 \label{b8b}
 \end{eqnarray}
 In the new variables the effective action can be again divided as in
 (\ref{b9a}), but with
 \begin{eqnarray}
 I_0&=&\int dx\lbrack p_\alpha{\partial v_\alpha\over
 \partial t}-\nu\, p_\alpha \nabla^2 w_\alpha\rbrack
 \nonumber \\
 &&+{i\over   2}\int p_\alpha (x){\cal D}_{\alpha\beta}(x-y)
 p_\beta(y)\,dx\,dy\,,
 \label{b10b}  \\
 I_{\rm int}&=&\int dx\,p_\alpha(x)w_\beta(x)\nabla_\beta
 w_\alpha(x)\ .
 \label{b11b}
 \end{eqnarray}
 Here ${\cal D}_{\alpha\beta}(x-y)$ is the correlation function of
 the random force $\mbox{\boldmath$\phi$}$ in Eq.(\ref{b7}) which is
 again assumed to be Gaussian (note that the initial force ${\rm f}$
 is not Gaussian now, but this is of no consequence).  In this formalism
 $\int D {\bf v} D{\bf p}$ is a functional integral over all possible
 realizations of the incompressible velocity field ${\bf  v}$ and the
 incompressible auxiliary vector field ${\bf p}$ at all points of space and
 time.  Note that in this formalism the auxiliary field ${\bf p}$
 multiplies the equation of motion for ${\bf v}$ and that because of the
 restriction of incompressibility we do not have to display the transverse
 projector in the effective action.  The quantity $Z$ in Eq.(\ref{b8b}) is
 a normalization constant,
 \begin{equation}
 Z=\int D{\bf v} D {\bf p}~\exp\lbrace iI\rbrace
 \label{b12}
 \end{equation}
 which is equal to $1$. The proof of this\cite{93KL} is based on the
 expansion of the $\exp\lbrace iI\rbrace$. The first term of this expansion
 is $1$. The diagrammatic representation of the next terms will have
 closed loops made of Green's functions; all these diagrams  are
 zero because of causality. Upon expanding in powers of  $I_{\rm int}$ we
 get from (\ref{b8b}) a theory that contains summations of infinite terms
 compared to the theory with ${\bf u}({\bf r},t)$ as the variable in the
 path integral. To see this note that (\ref{a2}),(\ref{a3}) can be
 rewritten as
 \begin{equation}
 {\bf u}( {\bf  r},t) ={\bf v}(   {\bf r}_0\vert   {\bf
 r}-\int^{t}_{t_0}{\bf v}({\bf r}_0\vert {\bf r}_0,\tau)d\tau,t)\ .
 \label{b13}
 \end{equation}
 The RHS  of this equation can be expanded in a Taylor series. The
 inversion of this series for ${\bf v}$ in terms of ${\bf u}$ is again an
 infinite series.  We thus see that given powers of ${\bf v}$ contain
 infinite summations in terms of powers of ${\bf u}$, and vice versa. It is
 extremely difficult to find the explicit diagrammatic resummation of the
 series in ${\bf u}$ to get the series in ${\bf v}$. It is a great
 advantage of the path integral formulation  that such infinite
 resummations can be obtained by a simple change of  variables.  This
 resummation is referred here as the Belinicher-L'vov  renormalization. We
 reiterate that this renormalization is an exact  formal step, and its
 usefulness needs to be demonstrated. It is already obvious however  that
 although the original functional (\ref{b8a}) produces correlators which
 are not Galilean invariant, the new one (\ref{b8b}) is free of this
 problem.  This guarantees that the theory that is derived on the basis of
 this functional is Galilean invariant.

 An average of any product of the fields ${\bf w}$
 and ${\bf p}$ is computed as a functional derivative of ${\cal Z}({\bf
 l},{\bf m})$ with respect to ${\bf l}$ and ${\bf m}$ in the  appropriate
 order of differentiation, finally taking the limit $l,m\to 0$. For example
 \begin{eqnarray}
 F_{\alpha\beta}({\bf r}_0\vert {\bf r}, {\bf r}',t,t')
 &=&\left\langle w_\alpha ({\bf r}_0\vert  {\bf r},t)w_\beta ( {\bf r}_0
 \vert   {\bf r}',t')\right\rangle
 \label{b14} \\
 &=&{\delta^2 {\cal Z}({\bf l}, {\bf m})\over\delta l_\alpha({\bf r},t)
 \delta l_\beta( {\bf r}',t')}\vert_{l,m\rightarrow 0}\  .
 \nonumber
 \end{eqnarray}
 The perturbation theory is obtained by expanding (\ref{b8b}) in
 $I_{\rm int}$ which contains the fields ${\bf w}$ and ${\bf p}$.  Since
 $I_0$  is a quadratic form in the  fields, the natural objects which
 appear in the diagrammatic  expansions are the correlators $\left\langle
 {\bf w}{\bf w} \right\rangle$, $\left\langle {\bf w} {\bf p}\right\rangle$
 and $\left\langle {\bf p} {\bf p}\right\rangle$. The correlator
 $\left\langle {\bf p} {\bf p}\right\rangle$ is zero identically in this
 approach\cite{76Dom,76Jan,95LP-a}, whereas $\left\langle {\bf w} {\bf
p}\right\rangle$
 is the Green's function, or the  response to an additional external noise
 ${\bf h}({\bf r},t)$\cite{76Dom,76Jan,95LP-a}
 \begin{eqnarray}
 &&G_{\alpha\beta}({\bf r}_0\vert {\bf r}, {\bf r}',t,t')
 \label{b15} \\
 &=&i \left\langle w_\alpha(   {\bf r}_0\vert   {\bf r},t) p_\beta(   {\bf
 r}',t')\right\rangle={\delta \left\langle w_\alpha(   {\bf r}_0\vert {\bf
 r},t)\right\rangle\over\delta h_\beta(   {\bf r}',t')}\vert_{h\rightarrow
 0}\ .
 \nonumber
 \end{eqnarray}

 The first step in setting up the theory is the calculation of the bare
 Green's function. The bare Green's function $G^0_{\alpha\beta}$ is the
 response in ${\bf w}$ (\ref{b15}) without the nonlinearity in the equation
 of motion.  We can also introduce a Green's function $\tilde
 G^0_{\alpha\beta}$ which is the response of the field ${\bf v}$
 without nonlinearity. From the equation of motion (\ref{b7}) we see that
 $\tilde G^0_{\alpha\beta}$ satisfies the equation
 \begin{equation}
 ({\partial\over \partial t}-\nu\nabla^2)\tilde
 G^0_{\alpha\beta}({\bf r}, {\bf r}',t,t')=
 P_{\alpha\beta}({\bf r}-  {\bf r}')\delta (t-t').
 \label{b16}
 \end{equation}
 where the RHS stems from (\ref{b2}) with ${\bf a}({\bf r})=\hat e\delta
 ({\bf r}- {\bf r}')$, $\hat e$ is some unit vector.  It is clear that
 $G^0_{\alpha\beta}$ can be understood as the difference
 \begin{equation}
 G^0_{\alpha\beta}({\bf r}_0\vert {\bf r}, {\bf r}',t,t')
 =\tilde G^0_{\alpha\beta}({\bf r}, {\bf r}',t,t')-\tilde
 G^0_{\alpha\beta}({\bf r}_0, {\bf r}',t,t')\ .
 \label{b17}
 \end{equation}
 From Eqs.(\ref{b16}-\ref{b17}) we conclude that for $t>t'$ the Green's
 function satisfies the equation
 \begin{eqnarray}
 \Big[{\partial\over \partial t}
 &-&\nu\big(\nabla^2_0+\nabla^2\big)\Big]
 G^0_{\alpha\beta}({\bf r}_0
 \vert {\bf r}, {\bf r}',t,t')
 \label{b18} \\
 &=&G^0_{\alpha\beta}({\bf r}_0\vert {\bf r},{\bf r}',0^+)\delta (t-t')
 \nonumber
 \end{eqnarray}
 where
 \begin{equation}
 G^0_{\alpha\beta}({\bf r}_0\vert {\bf r}, {\bf r}',0^+)
 =P_{\alpha\beta}( {\bf r}- {\bf r}')- P_{\alpha\beta}({\bf r}_0-
 {\bf r}')\ .
 \label{b19}
 \end{equation}
 The notation $t=0^+$ means the limit $t\rightarrow 0$ from above. The
 functions $P_{\alpha\beta}({\bf r}- {\bf r}')$ are given explicitly in Eq.
 (\ref{b5}).  We recall that all Green's functions vanish for time $t<t'$
 due to causality.  Note that the Green's function at zero time is not
 proportional to a delta function in ${\bf r} -{\bf r}'$.  This  is a
 consequence of the condition of incompressibility, which means that the
 speed of sound is infinite, allowing information to propagate in zero
 time.
 To derive the equation for the dressed Green's function we note that the
 nonlinear part of the effective action has exactly the form of the
 Navier-Stokes nonlinearity, but in terms of ${\bf v}$ rather than ${\bf
 w}$.  Therefore the perturbation theory that this formalism produces is
 identical, in terms of graphical notation and in integral representation,
 to the standard perturbation theory that was developed in Refs.
 \cite{61Wyl,73MSR,76Dom,76Jan}.  The important difference is that
 the {\it objects} that appear in the theory are correlators of ${\bf w}$
 and responses of ${\bf w}$, and the {\it initial condition}
 $G^0_{\alpha\beta}({\bf r}_0\vert {\bf r}$, ${\bf r}',0^+)$ displayed in
 Eq. (\ref{b21}) is totally different from the initial condition in the
 Eulerian theory in terms of ${\bf u}$. In the Eulerian theory
 $G^0_{\alpha\beta}(0^+)$ has only  one projection operator
 $P_{\alpha\beta}({\bf r}- {\bf r}')$ in contrast to (\ref{b19}). This
 difference  will have far reaching consequences as we see later. At this
 point however we can use the standard procedure that results in the Dyson
 and Wyld equations for the line renormalized correlator and Green's
 functions. The Dyson equation for the Green's function reads
 \begin{eqnarray}
 \Big[{\partial\over \partial t}&-&\nu\big(\nabla^2_0+\nabla^2\big)\Big]
 G_{\alpha\beta}(   {\bf r}_0\vert {\bf r},   {\bf r}',t)=
 G^0_{\alpha\beta}(   {\bf r}_0\vert   {\bf r}, {\bf r}',0^+)\delta (t)
 \nonumber \\
 &+&\int d   {\bf r}_2G^0_{\alpha\delta}( {\bf r}_0  \vert   {\bf r},
 {\bf r}_2,0^+)\int d   {\bf r}_1\int^t_0 dt_1
 \nonumber \\
 &\times&
 \Sigma_{\delta\gamma} (   {\bf r}_0\vert   {\bf r}_2,   {\bf
 r}_1,t_1) G_{\gamma\beta}(   {\bf r}_0\vert   {\bf r}_1,   {\bf
 r}',t-t_1)\,,
 \label{b21}
 \end{eqnarray}
 where $t'$ was taken to be zero.  Note that an immediate consequence of
 Eq. (\ref{b21}) is that at short times $\lim_{t\to 0^+ }\lbrack
 G_{\alpha\beta}(   {\bf r}_0\vert   {\bf r}, {\bf r}',t)\rbrack
 =G^0_{\alpha\beta}(   {\bf r}_0\vert   {\bf r},   {\bf r}',0^+)$.

 The Wyld equation for the correlation function has the form
 \begin{eqnarray}
 && F_{\alpha\beta}({\bf r}_0\vert {\bf r}, {\bf r}',t)=\int
 d {\bf r}_1d {\bf r}_2\int^\infty_0dt_1dt_2G_{\alpha\gamma}
 ( {\bf r}_0\vert {\bf r},{\bf r}_1,t_1)
 \nonumber \\
 &\times &[ {\cal D}_{\gamma\delta}({\bf r}_1-{\bf r}_2,t-t_1+t_2)
 +\Phi_{\gamma\delta}(   {\bf r}_0\vert   {\bf
 r}_1,   {\bf r}_2,t-t_1+t_2)\rbrack
 \nonumber \\
 &\times& G_{\delta\beta}
 (   {\bf r}_0\vert   {\bf r}',   {\bf r}_2,t_2) \ .
 \label{b22}
 \end{eqnarray}
 In equation (\ref{b21}) the mass operator $\Sigma$ is related to the
 ``eddy viscosity" whereas in Eq.(\ref{b22}) the mass operator $\Phi$ is
 the renormalized ``nonlinear" noise which arises due to turbulent
 excitations. The diagrammatic representation for $\Sigma$ and $\Phi$ is
 well  known, and is reproduced in Fig.1.

 \section{Properties of the dressed propagators}
 In this section we present those asymptotic properties of the propagators
 which are used in the proof of locality in Sect.4. The reader who is not
 interested in the details of the derivation can just pause to examine Eqs.
 (\ref{c7}-\ref{c11b}) and then Eq. (\ref{c51}) before reading Sect.4.

 \subsection{
         Rules for reading diagrams, and their physical meaning} 
\begin{figure}
 \epsfxsize=6truecm
 \epsfbox{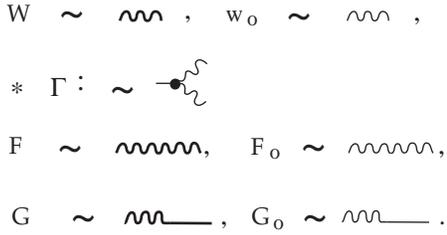}
 \caption{
 Graphical notation for the perturbation expansion.  The symbol  used are
 the following: short wavy lines stand for the fluid velocity  (think about
 waves). A thin wavy line stands for ${\bf u}_0$, whereas a bold  wavy line
 represents the full solution ${\bf u}({\bf r},t)$. A straight line stands
 for the field ${\bf p}({\bf r},t)$ that appear in the functional integral
 approach.  The Green's function, which is the response in the velocity to
 some force is made of a short wavy line and a short straight line
 representing the force.  Again this symbol appears in thin and bold
 variants. the former stands for the bare Green's function (2.20), and
 the bold for the dressed Green's function (2.18) The vertex  (3.1) is a
 fat dot with three tails.  One straight tail belongs to the Green's
 function, and two wavy tails stand for velocities.  A long wavy line will
 represent correlation functions of velocities.  Again, the thin and bold
 variants are bare and dressed correlators respectively. }
 \label{fig:fig1}
 \end{figure}

\begin{figure}
 \epsfxsize=8.6truecm
 \epsfbox{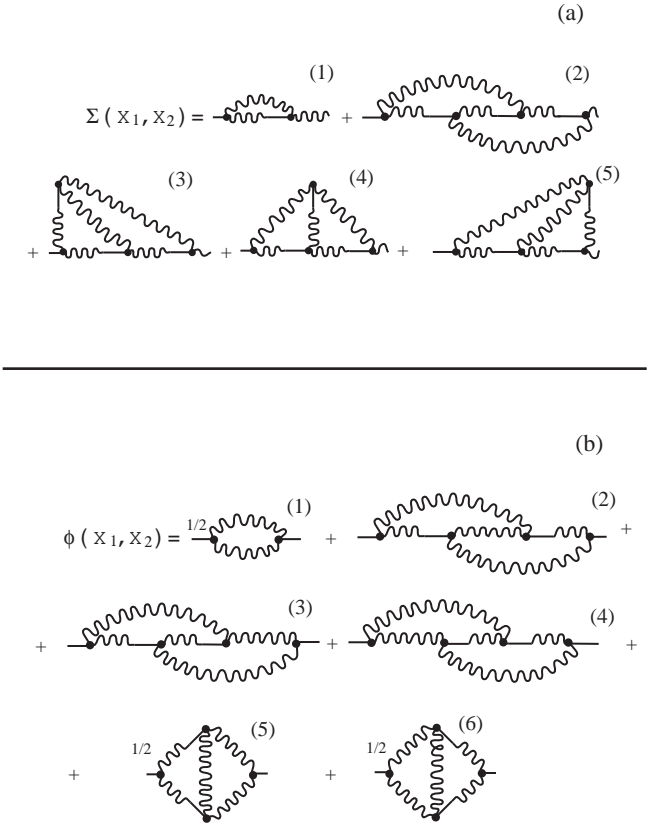}
 \caption{Diagrammatic representation of the  renormalized series
 expansion for the mass operators. (a) The mass   operator {$\Sigma$}
 of the Dyson  equation (2.23)  and (b) the mass operator {$\Phi$}
 of the Wyld equation (2.24).   }
 \label{fig:fig2}
 \end{figure}

The
 objects in the diagrams of Fig.\ref{fig:fig1} are the following: (i) the
 double correlator (\ref{b13}) of the BL-velocity differences ${\bf w}({\bf
 r}_0\vert {\bf r},t)$, which is denoted by a wavy line that carries the
  designation for the points $ {\bf r}_1,t_1$ and $ {\bf r}_2,t_2$, see
 Fig.\ref{fig:fig1}. (ii) The Green's function (\ref{b15}) which is
 represented half by a straight line that denotes the field ${\bf p}$, and
 half by  a wavy line that stands for ${\bf w}$. It also carries two
 space-time points, see Fig.\ref{fig:fig1}. Note that in contrast to the
 standard diagrammatic approach in the terms of the Eulerian velocity the
 objects also carry a designation for the special reference point ${\bf
 r}_0$.  (iii) The vertex $\Gamma_{\alpha\beta\gamma}$ is:
 \begin{equation}
 \Gamma_{\alpha\beta\gamma}=\delta_{\alpha\gamma}\nabla_\beta+\delta_
 {\alpha\beta}\nabla_\gamma\ .
 \label{c1}
 \end{equation}
 In diagrams this vertex is denoted by a bold dot that is the junction of
 one straight and two wavy lines. This vertex is identical to the one which
 appears in the standard theory in Eulerian variables\cite{61Wyl,73MSR}, and
 it is local in ${\bf r},t$ space. The three lines that join at the vertex
 carry at that  point the same space-time coordinate designation.  The
 differential operators in $\Gamma_{\alpha\beta\gamma}$ act only on the
 propagators that meet the vertex  with wavy lines. Every diagram appearing
 in the expansion for $\Sigma_{\alpha\beta}({\bf r}_0\vert {\bf r},t {\bf
 r}',t')$ has an entry with a straight tail associated with ${\bf r},t$ and
 an exit with a wavy tail associated with ${\bf r}',t'$.  Diagrams for
 $\Phi_{\alpha\beta}({\bf r}_0\vert {\bf r},t, {\bf r}',t')$ have two
 entries associated with straight tails carrying  the designation ${\bf
 r}$, $t$ and ${\bf r}'$ ,$ t'$ respectively, see Fig.\ref{fig:fig2}. Every
 other space-time point that appears in the inner vertices of the diagrams
 is integrated over.  For example the one-loop diagrams for $\Sigma$ and
 $\Phi$ read:
 \begin{eqnarray}
 \! \Sigma_{\alpha\beta}({\bf r}_0\vert {\bf r}, {\bf r}',t\!-\! t')&=&
 \! (\delta_{\alpha\epsilon}{\partial\over \partial r_\gamma}
 \! +\! \delta_{\alpha\gamma} {\partial \over \partial r_\epsilon})
 G_{\epsilon\rho}({\bf r}_0\vert  {\bf r}, {\bf r}',\! t-\! t')
 \nonumber \\
 \times (\delta_{\rho\beta}{\partial\over \partial
 r'_\delta}+\delta_{\rho\delta} { \partial\over \partial
 r'_\beta}  \!\!\!\!\!\!\!\! &&)
  F_{\gamma\delta}( {\bf r}_0\vert {\bf r}, {\bf r}',t-t')\, ,
 \label{c2} \\
 \! \Phi_{\alpha\beta}( {\bf r}_0\vert
 {\bf r}, {\bf r}',\! t-\! t') \! &=&\! {1\over 2}
 (\delta_{\alpha\epsilon}{ \partial\over \partial r_\gamma}
 +\delta_{\alpha\gamma} { \partial\over \partial r_\epsilon})
 (\delta_{\beta\rho}{ \partial\over \partial r'_\delta} \nonumber \\
 +\delta_{\beta\delta}{ \partial\over \partial r'_\rho})
 F_{\gamma\delta}({\bf r}_0\vert & {\bf r},& {\bf r}',t-t')
 F_{\epsilon\rho} ( {\bf r}_0\vert {\bf r},r',t-t')\ .
 \label{c3}
 \end{eqnarray}
 Note that the RHS of (\ref{c2}) is still an operator, since the operator $
 \partial/ \partial r'_\beta$  acts on the Green's function that appears to
 its right in the Dyson  equation (\ref{b21}). In contrast,
 $\Phi_{\alpha\beta}({\bf r}_0\vert {\bf r},t, {\bf r}',t)$ is a function,
 since the operators act only on the correlators that we see in
 Eq.(\ref{c3}). At the  one-loop order there are no inner vertices, and
 therefore (\ref{c2}) and  (\ref{c3}) do not have integrations. In general
 every diagram will have integrations over all intermediate vertices.

 The physical significance of the Dyson equation is that the dressed
 response is determined by hydrodynamic interactions involving intermediate
 points. For example, the response to forcing at ${\bf r}'$ has one direct,
 zero time contribution at ${\bf r}$. However, for any finite  time the
 response to forcing at ${\bf r}'$ is mediated by interactions at points $
 {\bf r}_1$ which via $\Sigma$ appear at point ${\bf r}$. In the one-loop
 approximation $\Sigma$ itself has a Green's function that mediates
 directly between ${\bf r}_1$ and ${\bf r}$.  In higher order contributions
 to $\Sigma$ there are sequential contributions  due to forcing at ${\bf
 r}'$ that are mediated by responses at various ${\bf r}_1, {\bf r}_2$ ...
 until ${\bf r}$ is reached.  $\Sigma$ represents the dressed response
 which is the sum  of all these sequential responses in the intermediate
 multiple sets of  points. The Green's functions that mediate intermediate
 points are  weighted by the correlators of velocity differences between
 these  points; if these correlators are relatively small, the contribution
 of the  Green's function to the total response is also small.

 The intuitive understanding of the Wyld equation is also straightforward.
 From the equation of motion (\ref{b7}), written schematically as
 \begin{equation}
 {\bf w} = {1\over\lbrack \partial/ \partial
 t-\nu\nabla^2\rbrack}\lbrace -\lbrack {\bf
 w}\cdot\mbox{\boldmath$\nabla$}\rbrack{\bf w}
 +\tilde{\mbox{\boldmath$\phi$}} \rbrace\,,
 \label{c4}
 \end{equation}
 we see that the
 nonlinear term $\lbrack {\bf w}\cdot \mbox{\boldmath$\nabla$}\rbrack {\bf
 w}$ is added to $\tilde{\mbox{\boldmath$\phi$}}$ and can be
 understood as an ``additional nonlinear noise" in the equation of  motion.
 It is natural to expect that the double correlator of the forcing  will
 have a nonlinear contribution of the type $\left\langle\lbrack {\bf
 w}({\bf r}_0\vert {\bf r}_1,t_1)\mbox{\boldmath$\nabla$}_1\rbrack
 {\bf w}( {\bf r}_0\vert {\bf r}_1,t_1)\lbrack {\bf w}({\bf r}_0\vert {\bf
 r}_2,t_2)\mbox{\boldmath$\nabla$}_2\rbrack {\bf w}({\bf r}_0\vert
 {\bf r}_2,t_2)\right\rangle$.  Indeed, the mass operator
 $\Phi_{\alpha\beta}( {\bf r}_0\vert {\bf r}, {\bf r}',t-t')$ can be
 written {\it exactly} as
 \begin{eqnarray}
 &&\Phi_{\alpha\beta}( {\bf
  r}_0\vert {\bf r},{\bf r}',t-t')
 \label{c6} \\
 &=& {\partial\over
 \partial r_{1\gamma}}{ \partial\over \partial r_{2\delta}}
 F_{\alpha\gamma\beta\delta}({\bf r}_0\vert {\bf r}_1, {\bf r}_1,
 {\bf r}_2, {\bf r}_2,t_1,t_1,t_2,t_2)\,,
 \nonumber
 \end{eqnarray}
 where $F_{\alpha\gamma\beta\delta}$ is the fourth order correlation of
 ${\bf w}$.  Thus Eq.(\ref{b22}) can be  understood as the result of
 squaring Eq.(\ref{c4}) and averaging, up to the  dressing of the Green's
 functions.  The content of the Wyld  diagrammatics is the dressing of the
 bare Green's function that appears in Eq.(\ref{c4}), and the
 representation of the fourth order $F_{\alpha\gamma\beta\delta}$ in terms
 of second order $F_{\alpha\beta}$. The dressing of the Green's function
 results from the cross terms between $\lbrack {\bf w}\cdot \mbox{\boldmath
 $\nabla$}\rbrace  {\bf w}$ and $\tilde{  \mbox{\boldmath$\phi$}}$ in
 (\ref{c4}).  Note that the  analytic form of the $\Phi_{\alpha\beta}$ in
  the one-loop order, Eq.(\ref{c3}) follows directly from the Gaussian
 decomposition of $F_{\alpha\gamma\beta\delta}$.

 \subsection{Zero time properties of the renormalized propagators}
 In the analysis of the asymptotic properties of the propagators it is
 convenient to translate the coordinate system such that ${\bf r}_0=0$.
 We  will denote the various quantities as $G_{\alpha\beta}(0\vert {\bf
 r},{\bf r}',t),$ $\Phi_{\alpha\beta}(0\vert {\bf r}, {\bf r}',t-t')$
 etc.  The asymptotic properties of the bare Green's function follow from
 Eq.(\ref{b19}), with the transverse projector given by (\ref{b5}).  Note
 that $G^0_{\alpha\beta}(0\vert {\bf r},{\bf r}',0^+)$ is not a
 symmetric function of ${\bf r}$ and ${\bf r}'$ (also the finite  time
 Green's function is not symmetric!). The physical meaning of
 $G^0_{\alpha\beta}(0\vert {\bf r},{\bf r}',0^+)$ is the zero time
 difference between two responses of ${\bf v}$ at the points ${\bf r}$ and
 $0$ to forcing at ${\bf r}'$. For $r'$  much smaller that r the  main
 contribution has to come from the response at the point ${\bf r}_0=0$:
 \begin{equation}
 G^0_{\alpha\beta}(0\vert {\bf r}, {\bf r}',0^+)\approx
 -P_{\alpha\beta}( {\bf r}') \sim{1\over r'^3}\ \  r'\ll r\ .
 \label{c7}
 \end{equation}
 The next order contribution is about $1/r^3$ which is much smaller.

 In studying the other limit $r\ll r'$ we note first that for $r$ zero (i.e
 $ {\bf r}= {\bf r}_0$ in arbitrary coordinates)  $G^0_{\alpha\beta}(0\vert
 0, {\bf r}',0^+)=0$ because of the exact cancellation of the two
 responses. For $r$ small compared to $r'$ we can expand the difference in
 responses in a Taylor series to find
 \begin{eqnarray}
 && G^0_{\alpha\beta}(0\vert {\bf r}, {\bf r}',0^+)
 \nonumber \\
 &\approx &
 -r_\gamma { \partial\over \partial r'_\gamma}P_{\alpha\beta}(
 {\bf r}')\sim {   {\bf r}\cdot   {\bf r}'\over r'^5\ }\ \ r \ll r'\ .
 \label{c8}
 \end{eqnarray}
 The last form demonstrates that in this limit $G^0_{\alpha\beta}(0\vert
 {\bf r}, {\bf r}',0^+)$ is an odd function of ${\bf r}$.

 The most important asymptotic properties of the correlator
 $F_{\alpha\beta}(0\vert   {\bf r},   {\bf r}',t)$ for our discussion are
 those that appear in its t=0 value.  We can find these properties for time
 $t=0$ by writing $F_{\alpha\beta}(0\vert   {\bf r},   {\bf r}',t)$
 identically as
 \begin{equation}
 F_{\alpha\beta}(0, {\bf r}, {\bf r}',0)
 = S_{\alpha\beta} ( {\bf r})+S_{\alpha\beta}(   {\bf r}')
 - S_{\alpha\beta}({\bf r}-   {\bf r}' )
  \label{c9}
 \end{equation}
 where
 \begin{equation}
  S_{\alpha\beta}(   {\bf r})\equiv
 \left\langle\lbrack u_\alpha(   {\bf r})-u_\alpha(0)\rbrack\lbrack
 u_\beta(   {\bf r})-u_\beta(0)\rbrack\right\rangle \ .
 \label{c10}
 \end{equation}
 The reason for (\ref{c10}) is that for $t=0$ the
 correlators of ${\bf v}$ coincide with  the correlators of ${\bf u}$
 as has been explained in Sec. 2. Denoting $S_2(r)\sim S_{\alpha\alpha}(
 {\bf r})$ we find that in the regime $r\ll r'$ Eq.(\ref{c9}) reads
 \begin{equation}
 F_{\alpha\beta}(0\vert {\bf r}, {\bf  r}',0)
 \sim S_2(r)  + r{ \partial S_2(r')\over \partial r'} \ .
 \label{c11a}
 \end{equation}
 Note that the function is symmetric in ${\bf  r}$, ${\bf r}'$, and in the
 opposite limit ${\bf r}\gg {\bf r}$' we simply have to exchange
 ${\bf r}$ and ${\bf r}'$ in (\ref{c11a}). For $t$ different  from
 zero the correlator $F_{\alpha\beta}(0\vert {\bf r}, {\bf r}',t)$ is
 always smaller (by the Cauchy-Schwarz inequality) than its value at $t=0$.
 This turns out to be all that  we need for our discussion.  The classical
 K41 estimate of (\ref{c11a}) is
 \begin{equation}
 F_{\alpha\beta}(0\vert {\bf r},   {\bf r}',0)
 \sim \bar\epsilon^{2/3}\left( r^{2/3} + {r\over  r'^{1/3}}\right) \ .
 \label{c11b}
 \end{equation}
 where $\bar\epsilon$ is the mean  energy dissipation per unit mass per
 unit time.

 \subsection{Decay of the Green's function in the short-time regime}
 For the purpose of the proof of locality to be developed in Section 4 we
 need to control the time integrals of the Green's function. To this end we
 will need to estimate the characteristic decay time of $G_{\alpha\beta}
 (0\vert {\bf r}, {\bf r}',t)$ at long times. This is done in Sec. 3.D.
 Here we begin by  estimating the characteristic decay time $\tau_0(r,r')$
 of $G_{\alpha\beta}(0\vert {\bf r}, {\bf r}',t)$ at short times $t \ll
 \tau_0(r,r')$. We will see that the first derivative of $G_{\alpha\beta}
 (0\vert {\bf r}, {\bf r}',t)$ vanishes at $t=0$.  Therefore we define
 $\tau_0( {\bf r}, {\bf r}')$ in the short time regime  with the help
 of the second derivative
 \begin{equation}
 {G_{\alpha\beta}(0\vert {\bf r}, {\bf r}',0^+)
 \over\tau^2_0( {\bf r}, {\bf r}')}
 \equiv{ \partial^2G_ {\alpha\beta}(0\vert {\bf r}, {\bf r}',t)
 \over \partial t^2}\vert_{t=0^+}\ .
 \label{c12}
 \end{equation} In principle $\tau_0( {\bf r},  {\bf r}')$ depends on the
 tensor indices $\alpha ,\beta$. We will not be interested in numerical
 factors and we therefore disregard this dependence. From the point of view
 of the scaling theory of turbulence, typical time scales depend on the
 involved length scales.  In K41 scaling the time scale $\tau (r)$ is
 proportional to $r^{2/3}$:
 \begin{equation}
 \tau (r)\sim\bar\epsilon^{1/3}r^{2/3}\ .
 \label{c13} \end{equation}
 We are  going to show that
 \begin{equation}
 \tau_0(r,r')\sim  \min\lbrace\tau (r),\ \tau (r')\rbrace = \tau
 (\min\lbrace r,r'\rbrace )\ .
 \label{c14}
 \end{equation}

 To compute the rate of change of $G_{\alpha\beta} (0\vert  {\bf r},  {\bf
 r}',t)$ we employ the  Dyson equation (\ref{b21}). For length scales $r$
 and $r'$ in the inertial range,  we can neglect the viscous term in
 (\ref{b21}).  The reason is that the  viscous contribution to the time
 derivative (\ref{c12}) is of the order of $\nu /r^2$, which is much
 smaller than $1/\tau (r)$. Initially we perform the evaluation of the mass
 operator $\Sigma$ on the RHS of (\ref{b21}) in the one-loop approximation,
 and then show that this evaluation is exact in the limit $t\rightarrow 0$.
 Substituting Eq.(\ref{c2}) into (\ref{b21}) and writing the one-loop
 result schematically, (disregarding tensor indices) we find for $t>0$
 \begin{eqnarray}
 {\partial\over  \partial t}G(0\vert   {\bf r},   {\bf r}',t)
 &=&\! \int\!  d {\bf r}_2G^0(0\vert {\bf r}, {\bf r}_2,0^+)
 \! \int \!  d{\bf r}_1\! \int^t_0 \! dt_1
 \label{c15} \\
 \times{\partial\over \partial  r_2}
 G(0\vert  {\bf r}_2,&\!\!{\bf r}_1,\!\!&t_1) { \partial\over  \partial
 r_1}\lbrack F(0\vert {\bf r}_2, {\bf r}_1,t_1) G(0\vert {\bf r}_1, {\bf
 r}', t-t_1)\rbrack) .
  \nonumber
 \end{eqnarray}
 Note that this schematic
 writing may be dangerous in that it obscures the possibility that some
 contributions might be odd under the  operations ${\bf r}_1\rightarrow -
 {\bf r}_1$ and/or $ {\bf r}_2\rightarrow -  {\bf r}_2$. To see this recall
 that the $  \partial / \partial {\bf  r}$ operations are odd whereas the
 rest of the integrand has even and odd contributions. Accordingly we will
 take the symmetry of the integrand into account when the result depends on
 it.

 The time arguments in $F$ and $G$  are bounded by $t$.  For $t \ll \tau
 (r)$ we can neglect the time dependence of these propagators on the RHS of
 (\ref{c15}), and use their zero time values. Performing the (trivial) time
  integral and differentiating with respect to time we find
 \begin{eqnarray}
 {\partial^2\over \partial t^2}\! &G&\!(0\vert {\bf r},{\bf
 r}',t)\!\approx \!\! \int  \! d   {\bf r}_2G^0(0\vert   {\bf r},   {\bf
 r}_2,0^+) \! \int \!\! d   {\bf r}_1{  \partial\over   \partial r_2}
 \label{c16}\\
 \times &G^0& (0\vert {\bf r}_2, {\bf r}_1,0^+)
 {\partial\over  \partial  r_1}
 \lbrack F(0\vert   {\bf r}_2, {\bf r}_1,0)G^0(0\vert  {\bf r}_1,
 {\bf r}',0^+)\rbrack \ .
 \nonumber
 \end{eqnarray}
 Next we note that the two-loop  order has two more time integrations that
 for small $t$ contribute to $ \partial G(0\vert {\bf r}, {\bf  r}',t)/
 \partial t$ starting at $O(t^3)$, and so to Eq.(\ref{c16})  starting at
 $O(t^2)$. The n'th loop contributes starting from $O(t^{2n-1})$ to $
 \partial G(0\vert  {\bf r},  {\bf r}',t)/ \partial t$. Thus we learn
 that for short times the expansion in the strength of interaction can be
 reformulated as an  expansion in powers of the time. In the limit
 $t\rightarrow 0$ the one-loop  approximation becomes exact. We can
 continue therefore to evaluate  our quantities for short times in the
 one-loop order with impunity.

 The analysis of the integral in (\ref{c16}) which we present next is
 quite cumbersome, due to the need to consider all the regimes of the
 relative sizes of $ {\bf r}, {\bf r}', {\bf r}_1$ and $r_2$. In
 performing this analysis we are  going to use the K41 estimate
 (\ref{c11b}) of the correlator. We stress  however that it will be shown
 that our conclusions are independent of  the precise value of the
 exponents as long as they are not too far from  the K41 exponents. This
 comment will be made quantitative later, when we understand the magnitude
 of the window of locality that is defined below. The result of our
 considerations is Eq. (\ref{c37a}).

 Denote the minimum and the maximum of $r$, $r'$ as
 \begin{equation}
 \lambda\equiv \min(r,r'), \Lambda\equiv
 \max(r,r')\ .
 \label{c17}
 \end{equation}
 It will be shown that the main contribution to the integral (\ref{c16})
 comes from the regime $r_1\sim r_2 \sim\lambda$. We begin by showing that
 the regimes in which $r_1$ or $r_2$ or both are  smaller than $\lambda$
 contribute negligibly to the RHS of (\ref{c16}).

 In the regime $r_1< \lambda <r_2$ the relevant part of the integral over
 $r_1$ in (\ref{c16}) is
 \begin{eqnarray}
 &&\!
 \int^\lambda_0 \!\! r^2_1  dr_1G^0(0\vert {\bf r}_2, {\bf r}_1,0^+)
 {\partial\over \partial r_1}\lbrack F(0\vert  {\bf r}_2,
   {\bf r}_1,0)G^0(0\vert  {\bf r}_1, {\bf r}',0^+)\rbrack
  \nonumber     \\
 &&\quad \sim
 {1\over r'^4}\int^\lambda_0 dr_1\left[ {1\over r_1^{1/3}}+{1\over
 r_2^{1/3}}\right]\ .
 \label{c18}
 \end{eqnarray}
 In the estimate of the LHS integral we used the evaluations  $G^0(0\vert
 {\bf r}_2, {\bf r}_1,0^+) \sim 1/r^3_1$ [cf. Eq.  (\ref{c7})], $ \partial/
 \partial r_1\sim 1/r_1,~F(0\vert   {\bf r}_2, {\bf r}_1,0) \sim
 [r^{2/3}_1+r_1/r_2^{1/3}]$ [cf. Eq.(\ref{c11b})] and $G^0(0\vert {\bf
 r}_1, {\bf r}',)^+)\sim r_1/r'^4$ [cf. Eq.(\ref{c8})].  We took here the
 next order term in $F$ because of the differentiation with respect to
 $r_2$ in (\ref{c16}). We see that in either case  the main contribution to
 (\ref{c18}) comes from the upper limit $r_1\sim\lambda$.

 Second we consider the regime $r_2<\lambda <r_1$. The relevant part of
 (\ref{c16}) now is
 \begin{eqnarray}
 \!\!\!\!\!\!\!\! &&
 \int^\lambda_0 \!\! r^2_2dr_2G^0(0\vert {\bf r},{\bf
 r}_2,0^+) {  \partial\over \partial r_2} \lbrack G^0(0\vert   {\bf r}_2,
 {\bf r}_1,0^+)F(0\vert {\bf r}_2, {\bf r}_1,0)\rbrack
 \nonumber \\
 && \quad \sim
 {1\over r^4_1}\int^\lambda_0{dr_2\over r_2^{1/3}}
 \label{c19}
 \end{eqnarray}
 where we used the evaluations $G^0(0\vert {\bf r}, {\bf r}_2,0^+)\sim
 1/r_2^3,~  \partial/ \partial r_2\sim 1/r_2,
 ~G^0(0\vert {\bf r}_2, {\bf r}_1,0^+)\sim r_2/r^4_1$ and $F(0\vert   {\bf
 r}_2,   {\bf r}_1,0)\sim r^{2/3}_2$. Again the contribution to  the RHS
 integral is from the upper limit.

 The regime $r_1,r_2 < \lambda$ is more delicate. Indeed, the same kind of
 evaluation of the integral in (\ref{c16}) will go as follows:  for
 $r_1\sim r_2 )$, $G^0(0\vert  {\bf r}, {\bf r}_2,0^+)\sim
 1/r^3_2,~G^0(0\vert   {\bf r}_2, {\bf r}_1,0^+)\sim 1/r^3_2$ , $F(0\vert
 {\bf r}_2,   {\bf r}_1,0)\sim r^{2/3}_2$  and the integral with respect to
 $r_1$ as $r_2^3$. Using these evaluations we find
 \begin{equation}
 {\partial ^2\over \partial t^2}G(0\vert {\bf r}, {\bf r}',t)\vert_{t=0}
 \sim
 r'^{-4}\int^\lambda_0\! dr_2/r_2^{4/3}\, ,\ {\rm wrong\,!}
 \label{c20}
 \end{equation}
 This integral indicates a divergence in the  regime $r_1\sim r_2 \ll
 \lambda$. This  apparent divergence is not really there;  it stems from
 disregarding  the oddness of the integral with respect to a change in sign
 of ${\bf r}_1$ and ${\bf r}_2$ simultaneously. The odd signature  that we
 missed is due to the ${\bf r}_2\cdot  {\bf r}'/ r'^5$ [cf.
 Eq.(\ref{c8})]. The rest of the integrand is even, and has been evaluated
 properly. Thus, looking at the next (even) order in the expansion of the
 last Green's function we get instead of $ {\bf r}_2\cdot {\bf  r}'/r'^5$ a
 term $( {\bf r}_2\cdot {\bf r}')^2/r'^7$.  Finally instead of  (\ref{c20})
 we get the following convergent integral
 \begin{equation}
 { \partial^2\over \partial t^2}G(0\vert   {\bf r},   {\bf r}',t)\vert_{t=0}
 \sim r'^{-5}  \int^\lambda_0dr_2/r_2^{1/3}\,,~~~\rm{right\,!}
 \label{c21}
 \end{equation}
 The other cases, namely $r_1\ll r_2<\lambda$ and $r_2\ll r_1<\lambda$ can
 be analyzed in the same manner with the same conclusion. In summary, we
 learn that the integrals   over $r_1$ and $r_2$ in (\ref{c16}) may take
 $\lambda$ as their lower limit.

 Next we show that the upper limit of the integrals over $   {\bf r}_1$ and
 $ {\bf r}_2$  are bounded by $\Lambda$.  In the regime $r_1 > \Lambda > r_2$
 the relevant part of (\ref{c16})  is similar to (\ref{c18}), and may be
 evaluated as
 \begin{eqnarray}
 &&\int^\infty_\Lambda r_1^2dr_1G^0(0\vert {\bf r}_2,  {\bf r}_1,0^+)
 \label{c22} \\
 &\times&
 { \partial\over \partial r_1}\lbrack  F(0\vert   {\bf  r}_2,   {\bf
 r}_1,0)G^0(0\vert   {\bf r}_1,   {\bf r}',0^+)\rbrack\sim{r^3_2\over
 r'^3}\int^\infty_\Lambda {dr_1\over r_1^{13/3}}\ .
 \nonumber
 \end{eqnarray}
 To get the RHS we used the evaluations $G^0(0\vert  {\bf r}_2,  {\bf
 r}_1,0^+)\sim   {\bf r}_2\cdot   {\bf r}_1 /  r_1^5+(   {\bf
 r}_2\cdot    {\bf r}_1)^2/r_1^7, F(0\vert   {\bf r}_2,   {\bf r}_1,0)\sim
 r^{2/3}_2+   {\bf r}_2\cdot   {\bf r}_1/r_1^{4/3}+  (   {\bf r}_2\cdot
 {\bf r}_1)^2/r_1^{10/3}$ and $G^0(0\vert   {\bf r}_1,   {\bf r}',0^+)\sim
 1/r'^3$. We need to consider higher order contributions  since the leading
 order in $F(0\vert   {\bf r}_2,   {\bf r}_1,0)$ drops in the $ \partial/
 \partial r_1$ differentiation  and the next order is odd.  The integral in
 (\ref{c22}) contributes in the  lower limit.

 In the regime  $r_2 > \Lambda > r_1$ the relevant integral looks like
 (\ref{c19}), and is evaluated as
 \begin{eqnarray}
 \! \int^\infty_\Lambda \!\!  r^2_2 dr_2 G^0 (0\vert   {\bf r},   {\bf
 r}_2,&{0^+}&){ \partial\over      \partial r_2}\lbrack G^0(0\vert   {\bf
 r}_2, {\bf r}_1,0^+)F(0\vert   {\bf r}_2,   {\bf r}_1,0)\rbrack
 \nonumber \\
 &\sim &{r^2\over r_1^{11/3}}\int^\infty_\Lambda{dr_2\over r^3_2}\ .
 \label{c23}
 \end{eqnarray}
 In this case we evaluated $G^0(0\vert  {\bf r}, {\bf r}_2,0^+)\sim   {\bf
   r}\cdot   {\bf r}_2/r^5_2+(   {\bf r}\cdot   {\bf r}_2)^2/r^7_2,~
 G^0(0\vert   {\bf r}_2,    {\bf r}_1,0^+)\sim 1/r^3_1,~F(0\vert   {\bf
 r}_2,   {\bf r}_1,0)\sim r^{2/3}_1+ {\bf r}_1\cdot   {\bf
    r}_2/r_1^{4/3}+(   {\bf r}_1\cdot   {\bf r}_2)^2/r_1^{10/3}$. We need
    the higher order terms for the same reasons as discussed  above. This
 integral also contributes at $\Lambda$.

 Finally  we need to consider the regime $r_1\sim r_2 > \Lambda$. The
 leading order evaluation is again zero by symmetry. Taking into account
 the next order contribution we evaluate
 \begin{equation}
  {   \partial^2\over      \partial t^2}G(0\vert   {\bf r},   {\bf
 r}',t)\vert_{t=0^+}\sim{r^2\over r'^3}\int ^\infty_\Lambda{dr_2\over
 r_2^{13/3}}\ .
  \label{c24}
 \end{equation}
 In summary, the contribution to  the integrals over $r_1$ and $r_2$ in
 (\ref{c16})  from regions much smaller than $\lambda$ and much larger than
 $\Lambda$ are  negligible.  This is the first example of the property that
 we refer to in  this paper as ``locality". This property allows us to
 estimate (\ref{c16}) as the definite integral
 \begin{eqnarray}
 &&\!\!\!\!\!\!\! {\partial^2\over \partial t^2} G(0\vert   {\bf r},   {\bf
 r}',t)\vert_{t={0^+}} \! \sim\!\! \int^\Lambda_\lambda \!\!\!
 dr_2r^2_2G^0( 0\vert {\bf r},   {\bf r}_2,0^+) \! \int^\Lambda_\lambda
 \!\!\! dr_1r^2_1 { \partial\over \partial r_2}
 \nonumber\\
 \!\!\! &\times& \! G^0 (0\vert   {\bf r}_2,
 {\bf r}_1,0^+){ \partial\over \partial r_1}\lbrack F(0\vert   {\bf r}_2,
 {\bf r}_1,0) G^0(0\vert {\bf r}_1,   {\bf r}',0^+)\rbrack .
 \! \label{c25}
 \end{eqnarray}
  The integral (\ref{c25}) offers us an easy evaluation of the
 characteristic time $\tau (r,r')$ when r and r' are of the same order,
 i.e.  when $\lambda=O(\Lambda)$. In this case we simply power count and
 find that
 \begin{equation}
 { \partial^2\over \partial t^2}  G(0\vert   {\bf r},  {\bf
 r}',t)\vert_{t=0^+} \sim {1\over r^{13/3}}\ .
 \label{c26}
 \end{equation}
 Comparing this result with Eq.(\ref{c12}) shows that
 \begin{equation}
 \tau_0(r,r)\sim \bar\epsilon^{-1/3}r^{2/3}\,,
 \label{c27}
 \end{equation}
 which is identical to the Kolmogorov time $\tau (r)$ given by (\ref{c13}).
 We note that this finding is far from trivial; it requires the proof of
 locality.  Once locality is shown, the analytical result must agree with
 dimensional analysis as we indeed find. It is important to understand
 however that Eq.(\ref{c25}) offers more than can be achieved by
 dimensional analysis even with locality. For example we can discuss
 now $\tau_0(r,r')$ when $r\gg r'$ and $r\ll r'$. Since the ratio $r/r'$ is
 dimensionless,  dimensional analysis cannot be employed to estimate the
 characteristic time $\tau (r,r')$ in such cases.

 The first case to be discussed is $r\ll r'$. We use Eq.(\ref{c25}) with
 $\lambda=r$ and $\Lambda=r'$. The first and last Green's functions in
 (\ref{c25}) are evaluated as $G^0(0\vert   {\bf r},   {\bf r}_2,0^+)\sim
 r/r^4_2$ and $G^0(0\vert   {\bf r}_1,   {\bf r}',0^+)\sim r_1/r'^4$. This
 leads to the integral
 \begin{eqnarray}
 &&{ \partial^2\over   \partial   t^2}G(0\vert   {\bf r},   {\bf
 r}',t)\vert_{t=0^+}
 \label{c28} \\
 & \sim & {r\over r'^4}\! \int^{r'}_r\!\! {dr_2\over
 r^2_2}\!\int^{r'}_r\! \! dr_1r^2_1{      \partial\over      \partial
 r_2}G^0(0\vert   {\bf  r}_2,   {\bf r}_1,0^+)F(0\vert   {\bf r}_2,{\bf
 r}_1,0) .
 \nonumber
 \end{eqnarray}
 We break now the integration over  $r_1$ into the two domains $(r,r_2)$
 and $(r_2,r')$. In the first domain the $r_1$ integrand is evaluated as
 $1/r_2^{4/3}$, and as $r_2^{2/3}/r_1^2$ in the second domain. In both
 domains therefore the main contribution comes for the regime $r_1 \sim
 r_2$.  Thus the $r_1$ integral is evaluated as $1/r_2^{1/3}$. With this we
 can see that the  $r_2$  integral in (\ref{c28}) contributes mainly at
 $r_2 \sim r$, and can be evaluated as $1/r^{4/3}$.  Finally
 \begin{equation}
 {   \partial^2\over   \partial t^2}G(0\vert   {\bf r}, {\bf
 r}',t)\vert_{t=0^+}\sim{1\over r'^4r^{1/3}},~~~~~~r\ll r'\ .
 \label{c29}
 \end{equation}
 Remembering that in this regime  $G^0(0\vert {\bf r},   {\bf r}',0^+)\sim
 r/r'^4$ we conclude with (\ref{c12}) that
 \begin{equation}
 \tau  (r,r')\sim(\bar\epsilon)^{-1/3}r^{2/3}\sim\tau(r),~~~~~~~~r\ll r'\ .
 \label{c30}
 \end{equation}

 In the opposite limiting case, i.e. $r\gg r'$, Eq.(\ref{c25}) is employed
 with $\lambda =r'$ and $\Lambda =r$. After the evaluation of the first and
 last Green's functions  the integral in (\ref{c25}) reads:
 \begin{eqnarray}
 &&{  \partial^2\over  \partial t^2}G(0\vert   {\bf r},   {\bf r}',t)
 \vert_{t=0^+}\sim  r'^{-3}\int^r_{r'}{dr_2\over r_2}
 \label{c31}  \\
 &\times& \int^r_{r'}dr_1r_1^2{  \partial\over  \partial
 r_2}G^0(0\vert   {\bf r}_2,   {\bf r}_1,0^+){ \partial\over  \partial
 r_1}F(0\vert   {\bf r}_2,   {\bf r}_1,0)\ .
 \nonumber
 \end{eqnarray}
 We split again the  $r_1$ integral into two domains of integration
 $(r',r_2)$  and $(r_2,r)$. In the first domain we evaluate $G^0(0\vert
 {\bf r}_2,   {\bf r}_1,0^+)\sim 1/r^3_1$, and $      \partial^2\lbrack
 F(0\vert   {\bf r}_2,   {\bf r}_1,0)] /      \partial r_1      \partial
 r_2\sim 1/r_2^{4/3}$. Substituting in the $r_1$ integral we  evaluate it as
 $\ln(r_2/r')/r_2^{4/3}$.  Evaluating then the $r_2$ integral we find that it
 contributes at $r_2\sim r'$. We end up with
 \begin{equation}
 {\partial^2\over      \partial   t^2}G(0\vert   {\bf r},   {\bf
 r}',t)\vert_{t=0^+}\sim {1\over r'^{13/3}}\ .
 \label{c32}
 \end{equation}
 In   the second domain with $r_1>r_2$ we have to take into account two
 terms in the evaluation of $G^0(0\vert   {\bf r}_2,   {\bf r}_1,0^+)$,
 since the contribution of the first one will vanish by symmetry
 $
 G^0(0\vert   {\bf r}_2,   {\bf r}_1,0^+)\sim    {\bf  r}_2\cdot   {\bf
 r}_1 / r^5_1+(   {\bf r}_2\cdot   {\bf r}_1)^2/ r^7_1
 $.
  Evaluating $ \partial F(0\vert  {\bf r}_2,   {\bf r}_1,0)/   \partial
  r_1\sim r_2/r_1^{4/3}$ we end up with the evaluation of the $r_1$
 integrand as $r_2^2/r_1^{13/3}$. The $r_1$ integral contributes at $r_1
 \sim r_2$ and is evaluated as $1/r_2^{7/3}$.  The $r_2$ integral
 contributes at $r'$, and can be evaluated as $1/r'^{4/3}$. This means that
 the two domains of integration lead to the same evaluation (\ref{c32})
 which is therefore the final evaluation in this regime of $r\gg r'$.
 Remembering that here $G^0(0\vert {\bf r},   {\bf r}',0^+)\sim 1/r'^3$ we
 end up, using (\ref{c12}) with
 \begin{equation}
 \tau_0(r,r')\sim(\bar\epsilon)^{-1/3}r'^{2/3}\sim\tau(r'), r'\ll r\ .
 \label{c34}
 \end{equation}

 The conclusion from Eqs.(\ref{c30}) and (\ref{c34}) is that in the short
 time regime the typical decay time of the Green's function $\tau (r,r')$ is
 always of the order of the smaller turn over times, as written in
 (\ref{c14}). This conclusion is intuitively quite reasonable.

 It is very important to notice that we have  proven  here locality in a
 stronger sense than the one invoked in (\ref{c25}). It was shown that both
 the $r_1$ and the $r_2$ integrals take their main contribution from the
 regime $r_1 \sim r_2 \sim\lambda = \min(r,r')$. Although we used the
 Kolmogorov scaling  for the evaluation of the integral, the result does
 not depend on that.  We have a wide window of locality. We can change the
 scaling exponents of $F(0\vert   {\bf r}_2,   {\bf r}_1,0)$ and/or of
 $\tau (r)$ by a small number and still all  our integrals will continue to
 be dominated by the same domains $r_1\sim r_2\sim\lambda$.

 In fact, we can derive the result (\ref{c14}) for the typical time scale
 without invoking any numerical value for the scaling exponents.
 Assume that all the integrals in (\ref{c25}) are local in the sense that
 the largest contribution comes  from the region of integration $r_1\sim
 r_2\sim\lambda =min(r,r')$. Under this assumption we can evaluate
 $G^0(0\vert   {\bf r}_1,   {\bf r}_2,0^+)\sim   1/\lambda^3,~F(0\vert
 {\bf r}_2,   {\bf r}_1,0)\sim S_2(\lambda)$ and $ \partial / \partial r_1
 \sim  \partial /  \partial r_2\sim \lambda^{-1}$.  Eq. (\ref{c25}) takes
 the form
 \begin{equation}
 {      \partial^2\over \partial t^2}G(0\vert
 {\bf r},   {\bf r}',t)\vert_{t=0^+}\sim \lambda G^0(0\vert r,\lambda
 ,0^+)S_2(\lambda )G^0(0\vert\lambda ,r',0^+)\ .
 \label{c35}
 \end{equation}
 Using Eq. (\ref{c12}) we find
 \begin{equation}
 \tau^2_0(r,r')\sim{G(0\vert   {\bf r},   {\bf r}',0^+)\over
 \lambda G^0(0\vert r,\lambda ,0^+)S_2(\lambda )G^0(0\vert \lambda
 ,r',0^+)}\ .
 \label{c36}
 \end{equation}
 Consider now the limiting case $r\ll r'$.  Now $\lambda =r$ and $\Lambda
 =r'$, and we can estimate $G(0\vert   {\bf r},   {\bf r}',0^+)\sim\lambda
 /\Lambda^4,~G^0(0\vert r,\lambda ,0^+)\sim 1/\lambda^3$  and $G^0(0\vert
 \lambda ,r',0^+)\sim\lambda /\Lambda ^4$.  Finally we get again
 (\ref{c14}) but with a more general expression for $\tau (\lambda )$:
 \begin{eqnarray}
 \tau_0(r,r')&=&\tau (\lambda )\ ,
 \label{c37a}\\
 \tau (\lambda )&\approx& {\lambda\over\sqrt{S_2 (\lambda)}}\ .
 \label{c37b}
 \end{eqnarray}
  In the opposite case in which    $\Lambda =r\gg r'=\lambda$ all the
  Green's functions are evaluated as $1/\lambda^3$ and we regain
  (\ref{c37a}). Thus (\ref{c37a}) is the final result  for the
 characteristic time scale. This time scale is recognized as the typical
 turnover time of eddies of size $\lambda$, having rms velocity of
 $\sqrt{S_2(\lambda )}$.  Clearly for K41 scaling we recapture the scaling
 law $\tau(\lambda ) \sim\lambda^{2/3}$, but in  general our result is that
 the decay time of the Green's function $G(0\vert   {\bf r},   {\bf r}',t)$
 is the turn over time of eddies of scale $\lambda=\min(r,r')$,
 independently of the exponent. Eq.(\ref{c37b}) is one of the scaling
 relations between exponents of static and dynamic quantities that are
 used later to find the scaling solution of our theory.

 Note that the typical time scale $\tau (r)$ becomes the only relevant
 time scale of $G(0\vert {\bf r}, {\bf r}',t)$  when $r$ is about $r'$.
 In fact, when  $r\sim r'$ this  is also the typical time scale of the
 correlator $F(0\vert   {\bf r}, {\bf r}',t)$, of the mass operator $\Sigma
 (0\vert {\bf r},   {\bf r}',t)$, etc. The reason is that the problem
    contains no dimensional parameters, and therefore all the objects of
 the theory with the same spatial scale r must have the same characteristic
 time $\tau (r)$. This is {\it not} an assumption, but a consequence of the
 structure of the theory and the property of locality that is proven in
 full in Sec.4.

 \subsection{Characteristic decay time of the Green's function}
 In the previous section we evaluated the characteristic decay  time
 $\tau_0(r,r')$ of $G(0\vert   {\bf r},   {\bf r}',t)$ in the short time
 regime   $t\ll \tau_0(r,r')$. For the  proof of locality we need to
 evaluate time integrals over $G(0\vert   {\bf r},   {\bf r}',t)$, and for
 this purpose we need to estimate the global decay time $\tau (r,r')$
 defined as
 \begin{equation}
 \int^\infty_0dt~G(0\vert   {\bf r},   {\bf
 r}',t)=\tau(r,r')G^0(0\vert   {\bf r},   {\bf r}',0^+)\ .
 \label{c38}
 \end{equation}
  Of course this definition makes sense only if the integral converges; we
 prove this fact next. For $r$ and $r'$ of different orders we have no {\it
 a priori} reason to expect that $\tau (r,r')$  is of the order of
 $\tau_0(r,r')$.

 To prove the convergence of the integral in (\ref{c38}) we integrate
  Eq.(\ref{b21})  in time between $t=-\epsilon$ and infinity. For $r,\ r'$
  in the inertial  interval we neglect the viscous term for the same
 reasons as discussed in Section 3C. We find
  \begin{eqnarray}
 && \int^\infty_{-\epsilon}{      \partial\over      \partial
 t} G_{\alpha\beta}(0\vert   {\bf
 r},   {\bf r}',t)dt=G^0_{\alpha\beta}(0\vert   {\bf r},   {\bf
 r}',0^+)\nonumber\\
 &+&\int^\infty_{-\epsilon}dt\int d    {\bf r}_2G^0_{\alpha\delta}(0\vert
 {\bf r},   {\bf r}_2,0^+)\int d  {\bf r}_1\int^t_0dt_1
 \label{c39} \\
 &\times&
 \Sigma_{\delta\gamma} (0\vert   {\bf r}_2,   {\bf r}_1,t_1)G_{\gamma\beta}
 (0\vert   {\bf r}_1,   {\bf r}',t-t_1)\ .
 \nonumber
 \end{eqnarray}
 The Green's function vanishes for $t\rightarrow\infty$, and it is
 identically zero  for all negative times. Accordingly, the LHS is zero,
 and after changing the order of integration over time on the RHS we arrive
 at
 \begin{eqnarray}
 &&\!\!\!\!\!\!  G^0_{\alpha\beta}(0\vert   {\bf r},  {\bf r}',0^+)
 = -\int \! d {\bf r}_2G^0_{\alpha\delta}(0\vert {\bf r}, {\bf r}_2,0^+)
  \! \int \! d {\bf r}_1 \! \int^\infty_0 \! dt_1
  \nonumber\\
  &\times &   \Sigma_{\delta\gamma}(0\vert {\bf r}_2, {\bf
  r}_1,t_1)\int^\infty_0 dtG_{\alpha\beta} (0\vert   {\bf r}_1,   {\bf
 r}',t)\ .
 \label{c40}
 \end{eqnarray}
 Since the LHS is  finite, both time   integrals on the RHS converge. Thus
 the definition (\ref{c38}) is meaningful, and we proceed to estimate $\tau
 (r,r')$.

 To evaluate $\tau (r,r')$ we forget the tensor indices, and then begin
 the analysis by assuming that the main contribution to (\ref{c40}) comes
 from the regime $r_1\sim r_2$. Evaluating $\int d   {\bf r}_2\sim r^3_1$,
 and making use of (\ref{c38})  we rewrite (\ref{c40}) as
 \begin{eqnarray}
 G^0(0\vert   {\bf r},   {\bf r}',0^+)=\int   r^5_1dr_1G^0(0\vert   {\bf
 r},   {\bf r}_1,0^+)\int^\infty_0dt_1     &&
  \nonumber \\
  \times \Sigma (0\vert   {\bf r}_1,   {\bf
 r}_1,t_1) \tau(r_1,r')G^0(0\vert   {\bf r}_1,   {\bf r}',0^+)&&  \ .
 \label{c41}
 \end{eqnarray}
  Since we proved that both time integrals in (\ref{c40}) are finite, the
  integral $\int^\infty_0dt_1\Sigma (0\vert {\bf r}_1,   {\bf r}_1,t_1)$
 can be evaluated as $\tau (r_1)\Sigma (0\vert {\bf r}_1,   {\bf
 r}_1,t_1=0)$.  The time scale $\tau (r_1)$ is the same as the one used in
 the context of the  Green's function, because there is only one spatial
 scale available here.  The mass operator $\Sigma (0\vert {\bf r}_1,   {\bf
 r}_1,t_1=0)$ can be evaluated in the one-loop approximation (\ref{c2}) as
 $S_2(r)/r^5$.  Using the scaling relation (\ref{c37b}) this is equivalent
 to
 \begin{equation}
 \Sigma (0\vert   {\bf r}_1,   {\bf  r}_1,t_1=0)
 \sim{1\over r^3_1\tau^2(r_1)}\ .
 \label{c42}
 \end{equation}
 Thus  we can write
  \begin{eqnarray}
 && G^0(0\vert   {\bf r},   {\bf r}',0^+)
  \label{c43} \\
 &\sim& \int  r^2_1dr_1G^0(0\vert   {\bf r},   {\bf
 r}_1,0^+){\tau(r_1,r')\over\tau (r_1)}G^0(0\vert   {\bf r}_1, {\bf
 r}',0^+)\ .
 \nonumber
 \end{eqnarray}
 We are going to explore the implications of (\ref{c43}) on the allowed
 behavior of the ratio $\tau (r_1,r')/\tau (r_1)$. As discussed above, for
 $r_1\sim r'$ this ratio should be of $O(1)$. For $r_1 \gg r'$ but both r
 and $r'$ in the inertial interval we can expect a scaling behavior of the
 ratio
 \begin{equation}
 {\tau (r_1,r')\over\tau
 (r_1)}\sim\left\lbrack{r'\over r_1} \right\rbrack^\alpha,~~~~~~~~r_1\gg r'
 \label{c44}
 \end{equation}
 with some  scaling exponent $\alpha$.  Similarly, for $r_1 \ll r'$, we
 expect that
 \begin{equation}
 {\tau (r_1,r')\over\tau   (r_1)}\sim\left\lbrack{r'\over
 r_1}\right\rbrack^\beta,~~~~~~r_1\ll r'\ .
 \label{c45}
 \end{equation}
 Consider now (\ref{c43}) in the domain $(\Lambda ,\infty)$. The evaluation
 of the integrand is straightforward, and we write
 \begin{equation}
 G^0(0\vert   {\bf r},   {\bf r}',0^+)\sim\int^\infty_ \Lambda{dr_1\over
 r^2_1}\left\lbrack{r'\over
 r_1}\right\rbrack^\alpha\ .
 \label{c46}
 \end{equation}
 Convergence in the    upper limit, which is guaranteed by the fact that
 the LHS is finite, requires $\alpha >-1$. In the domain $(0,\lambda )$ we
 evaluate
 \begin{equation}
 G^0(0\vert   {\bf r},   {\bf r}',0^+)\sim\int^\lambda_0 dr_1\left
 \lbrack{r'\over  r_1}\right\rbrack^\beta\ .
 \label{c47}
 \end{equation}
 Convergence in the lower limit requires $\beta <1$. With these
 restrictions $(\alpha >-1$ and $\beta <1)$ the main contribution to the
 integral (\ref{c43}) comes from the domain $(\lambda ,\Lambda)$. For $r\gg
 r'$ we evaluate the LHS and RHS of (\ref{c43}) as
  \begin{equation}
 {1\over r'^3}\sim {1\over r'^3}\int^r_{r'}{dr_1\over r_1}\left\lbrack{r'\over
 r_1}\right\rbrack^\alpha\ .
 \label{c48}
 \end{equation}
 In order to satisfy this equation the integral must contribute in the
 lower limit.  Thus $\alpha$ must satisfy $\alpha > 0$. This is a stronger
 constraint than the one required by convergence.

 For $r\ll r'$ we evaluate the LHS and RHS of (\ref{c43}) as
 \begin{equation}
 {r\over r'^4}\sim{r\over r'^4}\int^{r'}_r{dr_1\over r_1}
 \left\lbrack{r'\over r_1}\right\rbrack^\beta\ .
 \label{c49}
 \end{equation}
 This integral must contribute in the upper limit in order to satisfy the
 equation. This requires $\beta <0$. Combining these results with
 (\ref{c44}) and (\ref{c45}) we reach the final conclusion that
 \begin{equation}
 \tau (r,r')/\tau (r)<\rm const
 \label{c50}
 \end{equation}
 with const = O(1) independently of the relation between $r$ and $r'$. In
 terms of our goal which was the estimate of the integral over the
 Green's function, we can write our final conclusion in the form
 \FL
 \begin{equation}
 \!\int^\infty_0 \! \! dt G_{\gamma\beta}(0\vert   {\bf r},   {\bf
 r}',t) \!<\! \tau (r)G^0_{\alpha\beta}(0\vert   {\bf r},   {\bf
 r}',0^+);\ {\rm always!}
 \label{c51}
 \end{equation}
 The meaning of    this is deep: this equation says that the characteristic
 time of decay of the Green's function $G_{\gamma\beta}(0\vert   {\bf r},
 {\bf r}',t)$ is always bounded by  the characteristic time of the first
 coordinate (${\bf r}$) which is the  coordinate designating the difference
 between the two responses at ${\bf r}$ and $0$. In the coordinate  system
 with ${\bf r}_0\ne 0$ the decay of the Green's  function is bounded by the
 time scale of $\vert {\bf r}-   {\bf r}_0\vert$.

 We remember that up to now we considered the local   contribution $r_1\sim
 r_2$. There are good reasons to expect that this is the largest
 contribution. However, if it turns out that this is not the case, then
 Eq.(\ref{c40}) implies that the estimate of the characteristic time scale
 $\tau (r,r')$ is even smaller than the bound (\ref{c50}). Therefore the
 bound (\ref{c51}) can be used safely.
 \section{Proof of locality}
 In this section we prove that all the diagrams for the Green's function
 and correlators are dominated by interactions within a shell of a ball in
 physical space whose radius $r$ will be determined  later. The logic of
 the arguments is as follows. It is easy to evaluate an arbitrary diagram
 given that the coordinates ${\bf r}_j$ of all vertices lie within a ball
 of radius $r$. The aim is to show that any contributions in which any of
 these vertices lies outside the ball of radius $r$  is smaller that this
  ``local" contributions. We will begin with the case in which only one
  vertex lies outside, and analyze the different geometries in which this
  may occur. We than consider the case of a group of vertices which lie
  close to each other but outside the ball, and finally the case where
  there are two separated groups of vertices lying outside the ball. This
  can be than generalized to any number of groups. The fact that all our
  integrals converge in both the upper and the lower limits serves as a
  proof that there are no perturbative mechanisms to renormalize the K41
  exponents. This is the main point of this paper.

 \subsection{The analysis of outer  fragments of diagrams}

\begin{figure}
 \epsfxsize=8.6truecm
 \epsfbox{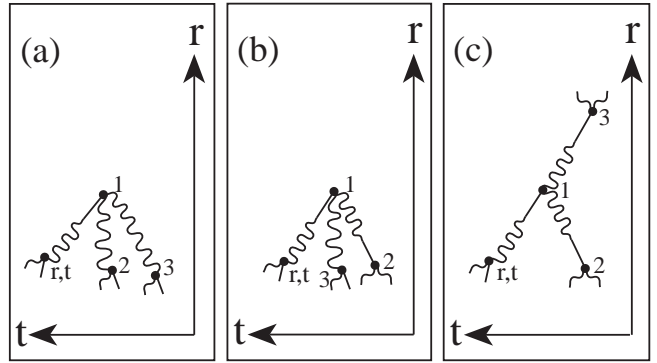}
 \caption{Three different arrangements of four adjacent
 vertices in a  fragment of a diagram, drawn in {${\bf r},t$} space. }
 \label{fig:fig3}
 \end{figure}

 We begin by proving that if the coordinate of one vertex, say $ {\bf
 r}_1$,  in an arbitrary diagram is removed far away from all the rest of
 the vertices, the effect is to reduce the evaluation of the diagram as an
 inverse power of $r_1$. Later we will use this result in our analysis of
 arbitrarily complicated diagrams. To facilitate the examination of the
 diagrams we will draw them in ${\bf r},t$ space, such that vertices at a
 larger value of r will be positioned above vertices at smaller values of
 $r$, and larger times will be to the {\it left} of smaller times. Consider
 now a fragment of an arbitrary diagram in which one vertex whose
 coordinate ${\bf r}_1$ is lying above all the other vertices in the
 diagram. In  Fig.\ref{fig:fig3} we show the three different possibilities
 of such a vertex at $ {\bf r}_1$  which is connected to another vertex at
 smaller ${\bf r}$ via a Green's  function. The relative positions in space
 and time of the four adjacent  vertices are clear from the pictures. We
 always take here $r_1>r$ and our  aim is to show that the major
 contribution to the integral over ${\bf r}_1$  comes from the region
 $r_1\approx r$.  Note that due to causality the vertex at ${\bf r}_1$ must
 be to the right of the vertex at ${\bf r}$, and any other Green's function
 that starts at $ {\bf r}_1$ has to point to the right. The wavy lines of
 the correlators can go either right or left, and we show only one
 representative example for each diagram that includes correlators.  Since
 we are going to estimate the correlators by their $t=0$ value, the time
 coordinates of the correlators do not matter. The three examples of
 Fig.\ref{fig:fig3} are denoted as $I_a,I_b$ and $I_c$.  The analytic
 expressions for these three examples are shown below. For simplicity we do
 not display tensor indices:
 \begin{eqnarray}
 I_a&\sim&\int^\infty_rd   {\bf
 r}_1\int^\infty_tdt_1{ \partial\over \partial r}G(0\vert   {\bf r},   {\bf
 r}_1,t_1-t)
 \label{d1} \\
 &\times &{\partial\over      \partial
 r_1}\lbrack F(0\vert   {\bf r}_1, {\bf r}_2,t_1-t_2)F(0\vert   {\bf r}_1,
 {\bf r}_3,t_1-t_3)\rbrack\ .
 \nonumber
 \end{eqnarray}
 We are going to
 bound  $I_a$ from above. Accordingly, we take the largest possible value
 of the correlators, which is the $t=0$ value. After that we can bound the
 integral over $t_1$ by using (\ref{c51}). We  use the symbols $\sim$  and
 $\propto$ in our evaluations, but we remember that we are actually
 estimating upper bounds; the quantities are {\it smaller} than what is
 displayed:
 \begin{equation}
 I_a\propto\tau   (r)\int^\infty_r{dr_1\over
 r^2_1}{ \partial\over \partial r_1}\lbrack F(0\vert   {\bf r}_1,   {\bf
 r}_2,0)F(0\vert   {\bf r}_1,   {\bf r}_3,0)\rbrack\ .
 \label{d2}
 \end{equation}
 We now evaluate
 \begin{equation}
 { \partial\over      \partial r_1}\lbrack F(0\vert   {\bf
 r}_1,   {\bf r}_2,0)F(0\vert   {\bf r}_1,{\bf r}_3,0)\rbrack\approx
 \bar\epsilon^{4/3}r_i^{2/3}r_j/r_1^{4/3}
 \label{d3}
 \end{equation}
 where $r_j$ is $\max\lbrace r_2,r_3\rbrace$, and $r_i=\min\lbrace
 r_2,r_3\rbrace$. Obviously, the integrals in  (\ref{d2}) contributes at
 $r_1\approx r$.

 The integral $I_b$ is
 \begin{eqnarray}
  I_b&\sim& \int d   {\bf r}_1\int^{t_2}_tdt_1{      \partial\over
   \partial r}G(0\vert   {\bf r},   {\bf r}_1,t_1-t)
  \label{d4} \\
 &\times&
 { \partial\over   \partial r_1}\lbrack G(0\vert   {\bf r}_1,   {\bf
 r}_2,t_1-t_2)F(0\vert {\bf r}_1,   {\bf r}_3,t_1-t_3)\rbrack\ .
 \nonumber
 \end{eqnarray}
  In   evaluating this integral we need to remember that in the generic
  diagrams the coordinate $t_2$ is integrated upon. The characteristic
 decay  time of the Green's function $G(0\vert   {\bf r}_1   {\bf
 r}_2,t_1-t_2)$ is bounded by $\tau (r_1)$.  However because this Green's
 function is connected to other propagators via vertex 2, the time
 restriction is at most $\tau(r_2)$ and not $\tau(r_1)$.  Thus we write
 \begin{equation}
  I_b\propto \tau(r)\int^\infty_r{dr_1\over r^2_1}{      \partial\over
  \partial r_1} \lbrack G(0\vert   {\bf r}_1,   {\bf
 r}_2,0)F(0\vert   {\bf r}_1,   {\bf  r}_3,0)\rbrack\ .
 \label{d5}
 \end{equation}
  We can evaluate now
 \begin{equation}
 {\partial\over      \partial r_1}\lbrack G(0\vert   {\bf r}_1,   {\bf
 r}_2,0)F(0\vert {\bf r}_1, {\bf r}_3,0)\rbrack\approx  \bar\epsilon^{2/3}
 {\hat {\bf r} _1\cdot   {\bf r}_2\over r^3_3r_1^{2/3}}\  .
 \label{d6}
 \end{equation}
 We see that the integral (\ref{d5}) is  local, contributing mostly at
 $r_1\sim r$.

 Next we consider the integral $I_c$:
 \begin{eqnarray}
 I_c&\sim& \int d {\bf r}_1\int^{t_i}_tdt_1{  \partial\over  \partial
 r}G(0\vert   {\bf r},   {\bf r}_1,t_1-t)
 \label{d7}   \\
 &\times & {\partial\over
 \partial r_1}\lbrack G(0\vert   {\bf r}_1,   {\bf r}_2,t_1-t_2)G(0\vert
 {\bf r}_1,   {\bf r}_3,t_1-t_3)\rbrack
 \nonumber
 \end{eqnarray}
 where $t_i = \min [ t_1,t_2 ]$.  In this case the integral
 $\int G[0\vert   {\bf r}_1,   {\bf r}_3,(t_1-t_3)] dt_1$ does contribute
 a factor bounded by $\tau (r_1)G(0\vert   {\bf r}_1,   {\bf r}_3,0^+)$.
 We thus need to take into account the factor $\tau (r_1)$ in the integrand
 and write
 \begin{equation}
 I_c\propto\tau  (r)\int^\infty_r{dr_1\over r^2_1}{      \partial\over
 \partial r_1}\lbrack\tau (r_1) G(0\vert   {\bf r}_1,   {\bf
 r}_2,0)G(0\vert   {\bf r}_1,{\bf r}_3,0)\rbrack\  .
 \label{d8}
 \end{equation}
 In the present case we use (\ref{c7}) and  (\ref{c8}) to write
 \begin{equation}
 { \partial\over \partial r_1}\lbrack  G(0\vert{\bf r}_1,{\bf r}_2,0)
 G(0\vert{\bf r}_1,{\bf r}_3,0)\rbrack\sim{\hat{ {\bf r}} _1\cdot\hat {
 {\bf r}}_j\over r^3_ir_1^{8/3}}
 \label{d9}
 \end{equation}
 where $\hat {\bf r} _j$   is a unit vector in the direction of $   {\bf
 r}_j$.  Obviously also $I_c$ is local.

 \subsection{Considerations involving arbitrary diagrams}
 We can consider now arbitrary diagrams appearing in the Dyson  equation.
 Every such diagram has a sequence of Green's functions connecting an entry
 at ${\bf r}$ to an exit at ${\bf r}'$ via intermediate points ${\bf r}_1,
 {\bf r}_2,...   {\bf r}_n$, see Fig.\ref{fig:fig2}, panel (a). We call
 this sequence of Green's functions the  ``principal path" of the diagram.
 From vertices along this principal path  there emanate either a double
 correlator or a sub-tree that begins  with a Green's function that serves
 as its root.

 Consider first the situation in which  $r'\leq r$. Our aim is to show that
 the main contribution to the diagram arises from the region of
 integration in which all the vertices are within the ball of locality of
 radius $O(r)$. Accordingly, we estimate first the mass operator $\Sigma$
 in the case that all the vertices are within the ball of locality. Using
 Eq.(\ref{c2}) the evaluation of $\Sigma$ within the one-loop approximation
 is straightforward:
 \begin{equation}
 \Sigma (r,r)\sim S_2(r)/r^5\ .
 \label{d10}
 \end{equation}
  The two-loop  diagrams contain two additional Green's functions which
  contribute a factor of $1/r^6$,  one correlator that contributes a factor
 of $S_2(r)$ [cf.(\ref{a6})], two vertices that contribute a factor of
 $1/r^2$, and two additional space-time integrations which give a factor of
 $r^6\tau (r)^2$.  Together all these additional factors give $S_2(r)\tau
 (r)^2/r^2$. Using the scaling relation (\ref{c23}) we conclude that this
 is $r^0$, and the two-loop diagrams are of the same order as the one-loop
 diagram. In fact, every order diagram in $\Sigma$ gives the same
 evaluation as (\ref{d10}).  We remark that this is a common property of
 scale-invariant solutions in strong coupling systems.

\begin{figure}
 \epsfxsize=8.6truecm
 \epsfbox{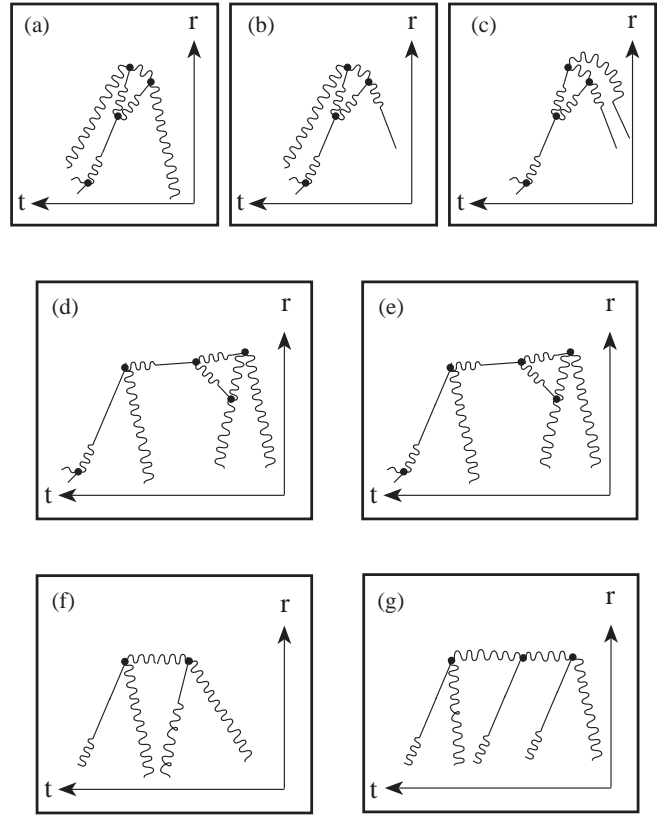}
 \vspace{.4cm}
 \caption{Fragments of diagrams which are used in the proof of
 locality. }
 \label{fig:fig4}
 \end{figure}

 Next consider a contribution in which the coordinates of all the  vertices
 in the diagram except one, say at a point $r_n$, lie within a ball of
 radius r. This case can be dealt with using the analysis of sec.4.1:  the
 situation locally must be one of the geometries (a), (b) or (c) in
 Fig.\ref{fig:fig3}.  But we have shown that in all these cases the main
 contribution comes from the region $r_n\approx r$. In fact the evaluation
 of the contribution will be diminished by a factor of $(r/r_n)^\alpha$
 where the exponent $\alpha$ takes the following values according to the
 local geometry:
 \begin{equation}
 \alpha =\cases{7/3~~(a)\cr
                  7/3~~(b)\cr
                  13/3~(c)\cr}
 \label{d11}
 \end{equation}
 A situation in which a number of vertices go out of the ball of radius $r$
 but do not have direct interactions between them is just a simple multiple
 of Eq.(\ref{d11}). Therefore the next situation to consider is that of a
 contribution with an arbitrary number of vertices outside the ball of
 radius $r$ with interactions. We take all of them to have $r_n$ of the
 same order, say of order~$\tilde r\gg r$. Such a blob of vertices can be
 connected to the ball of radius $r$ via three legs again as in the
 situation (a), (b) and (c), or via a larger number of legs, see
 Fig.\ref{fig:fig4}.
 First consider the decoration of the uppermost vertex as in
 Fig.\ref{fig:fig4}  a,b and
 c. The decorated vertex has two additional Green's functions, one
 additional correlator, and two space-time integrations. As we have learned
 such a combination scales like $r^0$. Notwithstanding, the estimates of
 the reduction in the importance of these contributions compared to
 (\ref{d10}) change from (\ref{d11}). The examples in Fig.\ref{fig:fig4}
 a, b, c are evaluated as
 \begin{equation}
 \Sigma\sim\lbrace S_2(r)/r^5\rbrace (r/\tilde  r)^2\ .
 \label{d12}
 \end{equation}
 The reasons for this evaluation are as  follows:  (i) The evaluation of
 the decorated vertex is $1/\tilde r$ (recall that the bare vertex is $
 \partial / \partial\tilde r$ which has  the same evaluation); (ii)  the
 Green's function connected via a straight  line to the decorated vertex is
 evaluated as $r/\tilde r^4$; (iii) the space integration gives $\tilde
 r^3$.  Lastly, the two additional legs in Fig.\ref{fig:fig4}, a have no
 $\tilde r$ dependence, and in toto we have (\ref{d12}). Note that in 5b
 and 5c we have one and two Green's functions respectively.  One could
 think that the time integrations over these Green's functions would
 introduce factors $\tau(\tilde r)$ and  $[\tau(\tilde r)]^2$ respectively.
 This is not so; the time  integration in the body of the diagram  is
 restricted  already by $\tau(r)$ because of the propagators in the body.
 Therefore the appropriate factors to use are $\tau (r)$ and $[\tau (r)]^2
 $ that do not change the dependence on $\tilde r$.

 Decorating the   vertex further (which is equivalent to looking at higher
 order contributions to the renormalized vertex) does not change the
 estimate (\ref{d12}). The scaling relation (\ref{c37b}) protects the
 evaluation of the contribution.  Therefore we consider now the situation
 in which the cluster of vertices at $\tilde r$ contains  an additional
 vertex that does not decorate the bare vertex, and which comes with an
 additional leg that connects the blob at $r$ (Fig\ref{fig:fig4} d and e).
 An additional vertex is worth $1/\tilde r$. The additional Green's
 function is worth $1/\tilde r^3$, and the additional space integration
 brings in~ $\tilde r^3$. The additional time integral is irrelevant since
 it is restricted by the shorter times in the body of the diagram.  Adding
 more legs connected to the blob always reduces the  contribution further.

\begin{figure}
 \epsfxsize=8.6truecm
 \epsfbox{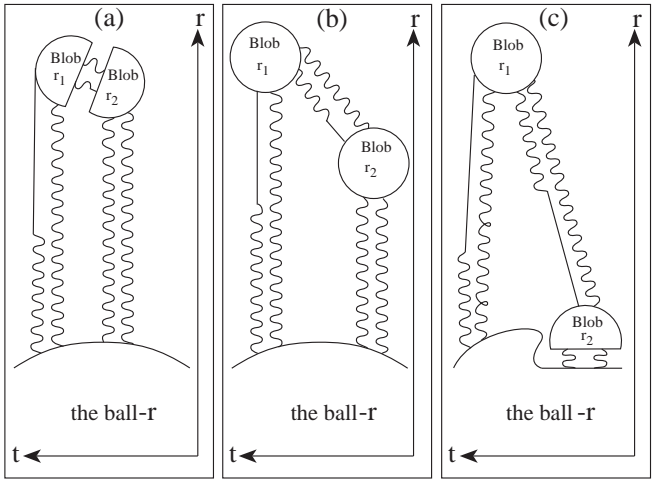}
 \vspace{.4cm}
 \caption{ Fragments of diagrams which are used in the global
 considerations in the  proof of locality.   }
 \label{fig:fig5}
 \end{figure}

 To complete the proof that the diagrams in the Dyson equation  are
 dominated by their local contribution we now consider the situation that
 two blobs of vertices are located outside the ball of locality, say one
 blob at $\tilde r_1$ and $\tilde r_2$, see Fig.\ref{fig:fig5}. There are
 two distinct possibilities:  (i) the reduction factor depends on $\tilde
 r_2$ like some power of $\tilde r_2$ and (ii) the reduction factor is
 independent of $\tilde r_2$. In case (i) the power may be either positive
 or negative.  If it is positive, then the main contribution  comes from
 $\tilde r_2 \sim \tilde r_1$, as shown in Fig.~\ref{fig:fig5}a. If the
 power is negative, the main contribution comes from $\tilde r_2\approx
 r$, as shown in Fig.~\ref{fig:fig5}c. Both situations have one blob
 outside the ball of locality, and  are covered by the discussion above.
 In case (ii) we will gain a logarithm $\log(\tilde r_1/\tilde r_2)$ that
 will multiply the power $(r/\tilde r_1)^\alpha$ that was  derived above
 for the single blob situation. This of course does not change the
 conclusion that the main contribution comes from the ball  of locality.

 The situation in which $r$ and $r'$ are of different orders of magnitude
 poses the question whether the ball of locality is of radius $r$ or $r'$.
 This question can be answered using the techniques developed above, but
 the natural framework to deal with this issue is to make use of the
 property of ``rigidity" which will be defined in paper II of this
 series\cite{95LP-b}. The result\cite{95LP-b} is that the ball of locality is
 always determined by the first coordinate in the case of the Green's
 function $G(0|{\bf r}, {\bf r}',t)$ and by the smaller coordinate in
 the case of the correlation function  $F(0|{\bf r}, {\bf r}',t)$.

 \subsection{Locality with respect to small intermediate
 coordinates and small separation distances}

 So far we have only considered the possibility that intermediate  points $
 {\bf r}_j$ go out of the ball of locality, and we have shown that this
 diminishes drastically the contributions to our diagrams. It is known that
 this is the dangerous divergence that appears in the Eulerian perturbation
 methods. Nevertheless we discuss here for completeness the possibility
 that intermediate points $   {\bf r}_j$ approach $   {\bf r}_0$, or the
 origin of  coordinates in our system that is centered at $   {\bf r}_0$.

 The discussion parallels to a large extent the steps that we took
 for the large distance case. Consider the situations of
 Fig.\ref{fig:fig3}, but now since $r_1$  is the smallest coordinate in the
 group, one should invert the coordinate system. This is like looking at
 the figures upside down, and mirrored across a line parallel to the ${\bf
 r}$ axis. (Causality always forces $G(0\vert   {\bf r},   {\bf
 r}_1,t_1-t)$ to point from left to right in our coordinates where $t$
 increases to the left). The main difference with the previous  calculation
 can be seen right away, say in the situation (a):
 \begin{eqnarray}
 {\rm (a)} &=&\tau (r)\int^r_0r^2_1dr_1{ \partial\over      \partial
  r}G(0\vert {\bf r},   {\bf r}_1,0^+)
  \label{d14}  \\
 &\times&
 { \partial\over      \partial
  r_1}\lbrack F(0\vert   {\bf r}_1,   {\bf r}_2,0)F(0\vert   {\bf r}_1,
 {\bf r}_3,0)\rbrack\ .
 \nonumber
 \end{eqnarray}
 Since the leading contribution   to $G(0\vert   {\bf r},   {\bf
 r}_1,0^+)$, i.e. $1/r_1^3$, disappears with the $      \partial/ \partial
 r$ differentiation, we take the next order which is $1/r^3$ [cf.
 Eq.(\ref{c7}) and the next line]. This factor does not influence the $r_1$
 integration, and we gain a factor of $r_1^4$ compared with the integration
 in the domain $(r,\infty)$. This difference now causes the integral to
 contribute at the upper limit, again at $r_1\approx r$. The same
 phenomenon  recurs in all the relevant integrals, and we leave the
 completion of the  proof, which is along the lines of the previous one, to
 the interested reader.

 Lastly, we need to consider also the situation in which two (or  more)
 coordinates approach each other, e.g $\vert   {\bf r}_{i,j}\vert =\vert
 {\bf r}_i-   {\bf r}_j\vert\ll r$. It is an  immediate consequence of our
 choice of BL variables that nothing  depends explicitly on this
 separation. Since we integrate over this separation from zero to $r$, the
 phase volume $r_{i,j}^2dr_{i,j}$ ensures that the  main contribution comes
 from $\vert   {\bf r}_{i,j}\vert\sim r$.

 In summary, we have shown that the exact resummation of the  Eulerian
 perturbation theory which results from the Belinicher-L'vov transformation
 leads to a new perturbation theory in which all the diagrams are local. We
 reiterate that ``locality" here means that the major contribution comes
 from intermediate coordinates that are in a shell of radius $r$. The width
 of this shell is also of the order of $r$. This was proven in the last
 paragraph when we showed that the contributions from the center of the
 ball of locality with $r_1\ll r$ are also negligible. If we measure
 separation distances on a logarithmic scale, as is appropriate in a scale
 invariant situation, we can state that the major contribution to the
 diagrams come from the narrow shell of the ball of locality.

 We can state these results in an intuitive fashion. With a  perturbation
 at a point ${\bf r}'$, from observing the differences of responses at
 points ${\bf r}$ and $ {\bf r}_0$, we learn from our diagrammatic
 expansion that the  coordinates of all the vertices [which designate the
 position of the hydrodynamic interactions $({\bf u}\cdot{\bbox{\nabla  }})
    {\bf u}$] that appear in any order of the  expansion must lie within a
 physical ball of fluid of radius $\vert   {\bf r}-   {\bf r}_0\vert$ in
 order to have a significant effect.  Also the double correlator is
 affected  by intermediate points in a finite ball of fluid of radius
 $\max(\vert   {\bf r}-   {\bf r}_0\vert,~\vert   {\bf r}'-   {\bf
 r}_0\vert)$. This is an analytic demonstration of the locality of the
 interactions  between turbulent fluctuations in physical space.
 \section{Discussion: the implications for scaling solutions}
\begin{figure}
 \epsfxsize=8.6truecm
 \epsfbox{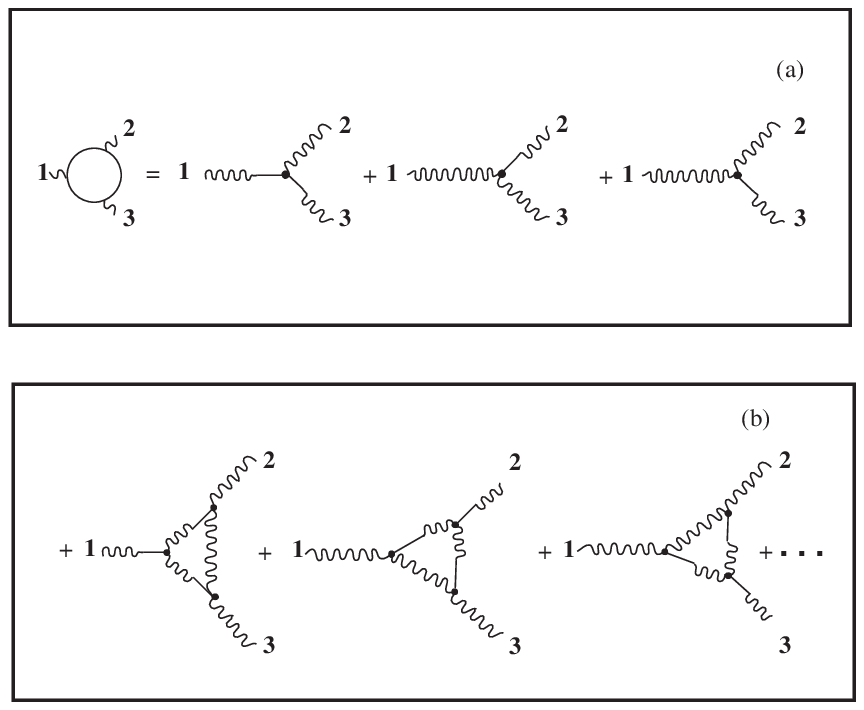}
 \vspace{.4cm}
 \caption{Diagrams in the expansion for the triple correlation
 function. (a)  All the first order contributions, (b) some of the third order
 contributions.   }
 \label{fig:fig6}
 \end{figure}

 We can examine now the implications of the fact that, order by order, our
 diagrammatic expansion is  local. First we show that K41 scaling is
 indeed a solution. Taking

 \begin{equation}
 S_2(r)\sim  r^{\zeta_2}\,,\quad\tau (r)\sim r^z\,,
 \label{e1}
 \end{equation}
 we see that Eq.(\ref{c37b}), which has been
 obtained solely from the property of locality, furnishes one scaling
 relation between $\zeta_2$ and z, i.e.
 \begin{equation} \zeta_2+2z=2\ .
 \label{e2} \end{equation}
 We can  easily find a second scaling relation
 by using our theory to calculate the 3-point correlation function
 \begin{eqnarray}
 && F_{\alpha\beta\gamma}({\bf r}_0\vert {\bf r}_1,  {\bf
 r}_2,   {\bf r}_3,t_1,t_2,t_3) \label{e3}  \\ &=&\left\langle w_\alpha(
 {\bf r}_0 \vert   {\bf r}_1,t_1)w_\beta (   {\bf r}_0\vert   {\bf
 r}_2,t_2)w_\gamma (   {\bf r}_0\vert   {\bf r}_3,t_3)\right\rangle\  .
 \nonumber
 \end{eqnarray}
 The   simultaneous correlation
 $F_{\alpha\beta\gamma}(   {\bf r}_0\vert {\bf r},   {\bf r},   {\bf
 r},t,t,t)$ is identical to the 3-rd order structure function of the
 Eulerian field, cf. Eq.(\ref{a6}), and it is well known that for small
 viscosity it scales (in the inertial range) like $r$.  The 3-point
 correlation function has a totally analogous theory to the one described
 above for the 2-point function, and a typical first order diagram for
 this quantity is shown in  Fig.\ref{fig:fig6}.  By repeating the analysis
 that led to the proof of locality we see that the main contribution to
 the integral $\int d   {\bf r}_1dt_1$ Fig.~7a comes from the region
 $r_1\sim r$ and $t_1\sim\tau (r)$ . Consequently we can evaluate $S_3(r)$
 in the first order as \begin{equation} S_3(r)\sim\tau  (r)\lbrack
 S_2(r)\rbrack^2/r\ .  \label{e4} \end{equation} The next order
 contributions to $F_{\alpha\beta\gamma}$ are nothing but the
 renormalization of the bare vertex, see for example Fig.~\ref{fig:fig6}b.
 One can prove following the ideas of Sec.  4 that all these diagrams are
 local.  The property of locality and the scaling relation (\ref{c37b}) or
 equivalently (\ref{e2}) imply that all the n'th order diagrams have the
 same evaluation as the first order diagram (\ref{e4}). Since $S_3(r)\sim
 r$, we find a second scaling relation \begin{equation} \lbrack
 S_2(r)\rbrack^2\tau (r)\sim r^2\ .  \label{e5} \end{equation} In terms of
 the scaling exponents this  is \begin{equation} 2\zeta_2+z=2\ .
 \label{e6} \end{equation} The solution of  (\ref{e2}) and (\ref{e5}) is
 $\zeta_2=2/3$, z=2/3, which are the standard K41 exponents.

 The usual mechanisms to renormalize exponents are connected to the
 divergence of the integrals appearing in higher order diagrams.  The
 cutoff lengths that are needed to control the divergence can be used to
 form dimensionless corrections to the scaling functions, and thus to
 renormalize the exponents. Since our diagrams are local, there are no
 typical scales which appear in this way, and we do not have a simple way
 to renormalize the scaling exponents of the structure functions. Indeed,
 if there is a renormalization of the exponents, it must stem from
 nonperturbative effects.  The study of these effects and the elucidation
 of multiscaling is the subject of papers II and III in this series.

 \acknowledgments
 The content of this paper owes a great deal to some persistent criticism
 of the ${\bf k}, \omega$ presentation by U. Frisch and R.H. Kraichnan.
 We are particularly grateful to R. H. Kraichnan for some deep discussions
 on these issues, and to Adrienne Fairhall, Daniel Segel and Stefan Thomae
 for important critical comments on the manuscript.  This work has been
 supported in part by the The Naftali and Anna Backenroth-Bronicki Fund
for Research in Chaos and Complexity,  Minerva Foundation, Munich,
Germany, the German Israeli Foundation and the Basic Research Fund of the
 Israeli Academy of Sciences. IP acknowledges the hospitality of the
 Mittag-Leffler Institute, Stockhlom, where a portion of this manuscript
 was written.

 \appendix

 \section{Relation between the correlation functions before and after the
 BL transformation}
 \label{sec:append}

 In this Appendix (which follows \cite{95LL-c}) we examine the relation
 between the BL transformed correlation functions and the correlation
 functions of the Eulerian fields. The aim of this appendix is to clarify
 how the stationarity and homogeneity of the original fields are reflected
 in the properties of the transformed fields. We introduced the BL velocity
 ${\bf v}$ by the expression (\ref{a2}):  ${\bf v}(t,{\bf r})={\bf
 u}(t,{\bf r}+{\bbox\rho})$ where ${\bf u}$ is the Eulerian velocity and
 ${\bbox\rho}(t)$ introduced by (\ref{a3}) is governed by the equation
 \begin{equation}
 \partial{\bbox\rho}/\partial t= {\bf u}(t, {\bf
 r}_0+{\bbox\rho})= {\bf v}(t, {\bf r}_0)\,, \quad
 {\bbox\rho}(t=t_0)={\bbox 0} \ .
 \label{ap1}
 \end{equation}
 Here $t_0$ is
 a marked time and ${\bf r}_0$ is a marked point which can be chosen
 arbitrary. Equation (\ref{ap1}) describes the Lagrangian trajectory ${\bf
 r}_0 + {\bbox \rho}$ of the liquid point starting from the point ${\bf
 r}_0$ at $t=t_0$.

 Let us introduce a designation  $\Psi$ for the product of the BL
 velocities taken in different spatial-time points,
 \FL
 \begin{eqnarray}
 && \Psi\{{\bbox\rho}\}=v(t_1,{\bf r}_1)
 v(t_2,{\bf r}_2) \dots v(t_n,{\bf r}_n)
 \label{ap2} \\
 &=& u\left[t_1,{\bf r}_1+{\bbox\rho}(t_1)\right]
 u\left[t_2,{\bf r}_1+{\bbox\rho}(t_2)\right]\dots
 u\left[t_n,{\bf r}_n+{\bbox\rho}(t_n)\right] .
 \nonumber
 \end{eqnarray}
 Next we transform the expression for the different-time correlation
 function of BL velocities, $\langle \Psi\{ {\bbox\rho}\} \rangle$, into
 the form which is convenient for utilizing the relation between ${\bf u}$
 and ${\bf v}$.  For this goal we will treat ${\bbox \rho}$ as an
 independent field taking into account the equation (\ref{ap1}) via the
 corresponding functional $\delta$-function. Namely $\langle \Psi\{
 {\bbox\rho} \}\rangle$ can be rewritten as
 \begin{equation}
 \int {\cal D}{\rho}\langle \delta\left\{\partial{\bbox\rho}/\partial t-
 {\bf u}(t, {\bf r}_0+{\bbox\rho})\right\} \Psi\{{\bbox\rho}\}\rangle
 \,, \label{ap8}
 \end{equation}
 where $\int{\cal D}{\rho}$ means the path integral over all functions
 ${\bbox\rho}(t)$. The functional $\delta$-function, $\delta\{\dots\}$  can
 be represented in the exponential form with the help of  auxiliary field
 ${\bbox\beta}(t)$. Then we find
 \begin{eqnarray}
 &&\langle\Psi\{{\bbox\rho}\}\rangle= \int{\cal D}\rho{\cal D}\beta
 \exp\left(i\int dt \, {\bbox\beta} \partial{\bbox\rho}/\partial t \right)
 \nonumber \\
 &\times&\left\langle \exp\left(-i\int dt \, {\bbox\beta} {\bf u}(t,{\bf
 r}_0 + {\bbox\rho}) \right) \Psi\{{\bbox\rho}\}\right\rangle \ .
 \label{ap3}
 \end{eqnarray}

 In the above formal transformations we omitted the functional
 determinant arising at passing to integration over $\rho$, $\beta$.
 To recognize the value of the determinant we should first discretize
 the problem introducing instead of (\ref{ap1}) a discrete scheme, e.g.
 \begin{eqnarray}
 && {\bbox\rho}(t_{n+1})-
 {\bbox\rho}(t_{n})-
 {\bf u}\left(t, {\bf r}_0+
 {\bbox\rho}(t_{n})\right) \Delta t =0
 \,, \nonumber \\
 && t_n=t_0+n \Delta t \,.
 \label{ap7}
 \end{eqnarray}
 Now the functional $\delta$-function in (\ref{ap8}) is the product
 of usual $\delta$-functions with arguments defined by the left-hand
 sides of the equation (\ref{ap7}), the product being taken for all
 numbers $n\ne 0$ (since in accordance with (\ref{ap1})
 ${\bbox\rho}(t_{0})={\bbox 0}$). The functional integration in
 (\ref{ap8}) is now the product of integrations over
 ${\bbox\rho}(t_{n})$, $n\neq 0$. We see that for $n>0$ the
 consequent integration over ${\bbox\rho}(t_{n})$ actually gives
 unity since
 $$ \int d {\bbox\rho}(t_{n+1})
 \delta \left( {\bbox\rho}(t_{n+1})-
 {\bbox\rho}(t_{n})- \dots \right) =1 \,. $$
 But when we go over $n$ in the negative direction the
 integrals over ${\bbox\rho}(t_{n})$ produce nontrivial
 factors associated with a dependence of {\bf u} on
 ${\bbox\rho}$. The product of these factors is just the
 functional determinant noted above.

 Usually this problem is avoided by putting $t_0$ into $-\infty$,
 after that the actual region of $t$ produces the unitary
 determinant. It is possible to do only if we deal with a
 situation as a stationary one which is not sensitive to initial
 conditions. Just this reason explains that the corresponding
 functional determinant can be regarded to be equal to unity in
 the path integral formulation of the Wyld-Martin-Rose-Siggia
 technique. But our problem is in proving time
 homogeneity of BL correlation functions, therefore $t_0$ is
 a finite time and we should consider times $t$ both before and
 after $t_0$. Fortunately in our case the incompressibility
 condition ensures the unitary of the determinant. The point is
 that for small $\Delta t$ in the approximation needed for us
 the integrals
 $$ \int d {\bbox\rho}(t_{n-1})
 \delta \left( {\bbox\rho}(t_{n})-
 {\bbox\rho}(t_{n-1})- \dots \right) =
 1-\Delta t {\bbox \nabla}{\bf u} $$
 are equal to unity due to the incompressibility condition.
 Thus we see that the condition is of the crucial importance for
 the Belinicher-L'vov scheme.

 Taking into account the definition (\ref{ap1}) we conclude
 that the expression in the angle brackets in (\ref{ap3})
 is an average of a combination of Eulerian velocities
 ${\bf u}\left(t,{\bf r}+{\bbox\rho}(t)\right)$ with a function
 ${\bbox\rho}(t)$.  This average is obviously translationally invariant.
 It allows one to shift all space arguments at a constant vector ${\bf R}$.
 Let us choose ${\bf R}={\bbox\rho}(\tau)$ where $\tau$ is some time. Then
 (\ref{ap3}) acquires the form
 \begin{eqnarray}
 &&\langle\Psi\{{\bbox\rho}\}\rangle
 = \int{\cal D}\rho^\prime  {\cal D}\beta \exp\left(i\int dt \, {\bbox\beta}
 \partial{\bbox\rho}^\prime/\partial t \right)
 \nonumber \\
 &\times&\left\langle \exp\left(-i\int dt \, {\bbox\beta}
 {\bf u}(t,{\bf r}_0+{\bbox\rho}^\prime) \right)
 \Psi\{{\bbox\rho}^\prime \}\right\rangle \,,
 \label{ap4}
 \end{eqnarray}
 where $ {\bbox\rho}^\prime(t)={\bbox\rho}(t)-{\bbox\rho}(\tau)$.  Note that
 ${\bbox\rho}^\prime(\tau)={\bbox 0}$.  Integration over $\rho,\beta$ in
 (\ref{ap4}) gives
 \begin{equation}
 \langle\Psi\{{\bbox\rho}\}\rangle=
 \langle\Psi\{{\bbox\rho}^\prime\}\rangle \,,
 \label{ap5}
 \end{equation}
 where ${\bbox\rho}^\prime(t)$ is governed by the same equation (\ref{ap1})
 as ${\bbox \rho}$ but with the initial condition ${\bbox \rho}^\prime
 (\tau)= {\bbox 0}$.

 The relation (\ref{ap5}) means that BL correlation functions are
 stationary in time since they do not actually depend on the marked time
 $\tau$. Using (\ref{ap5}) one can easily prove the coincidence of
 simultaneous correlation functions of BL and Eulerian velocities.  Namely,
 taking in (\ref{ap2}) all times $t_1, t_2, \dots$ to be equal to $t$ and
 choosing in (\ref{ap5}) $\tau=t$ we immediately find
 \begin{eqnarray}
 &&\langle v(t,{\bf r}_1)v(t,{\bf r}_2) \dots v(t,{\bf r}_n)\rangle
 \nonumber \\
 &=&\langle u(t,{\bf r}_1)u(t,{\bf r}_2) \dots u(t,{\bf r}_n)\rangle \,,
 \label{ap6}
 \end{eqnarray}
 since ${\bbox\rho}^\prime (t)=0$ in this case. Although we examined
 correlation functions of BL velocities, all arguments presented above can
 be applied to arbitrary fields e.g. pumping forces. It implies e.g. that
 BL Green's functions are stationary in time and that their simultaneous
 values coincide with ones of Eulerian Green's functions.


 \end{document}